\documentclass[twocolumn,english]{revtex4-2}
\usepackage[T1]{fontenc}
\usepackage[utf8]{inputenc}
\usepackage{amsmath}
\usepackage{esint}

\makeatletter
\usepackage{amsfonts}
\usepackage{graphicx}
\usepackage{braket}
\usepackage{hyperref}

\makeatother

\usepackage{babel}
\begin{document}
\title{Reduced density matrix and entanglement Hamiltonian for a free real
scalar field on an interval.}
\author{Mikhail A. Baranov}
\affiliation{Department of Theoretical Physics, University of Innsbruck, Technikerstr. 21a,
A-6060 Innsbruck, Austria}
\affiliation{Institute for quantum Optics and Quantum Information of the Austrian
Academy of Sciences, Technikerstr. 21a, A-6020 Innsbruck, Austria}
\date{09 May 2026}
\begin{abstract}
An exact result for the reduced density matrix on a finite interval
for a $1+1$ dimensional free real scalar field in the ground state
is presented. In the massless case, an explicitly performed Williamson
decomposition of the appearing kernels is used to reproduce the known
result for the entanglement (modular) Hamiltonian, as well as to find
the leading ($\sim1/\ln M$) non-local corrections for a small mass
$M$ of the field. In the opposite case $M\to\infty$ , it is argued
that, up to terms $O(M^{-1/2})$, the entanglement Hamiltonian is
local with the density being the Hamiltonian density spatially modulated
by a triangular-shape function.
\end{abstract}
\maketitle

\section{Introduction}

For a quantum system in the ground or in any other state, a reduced
density matrix of its subsystem is an important source of information
about the entanglement properties of the state, which manifest its
complexity. This holds for field theories and many-body systems, in
the continuum and on a lattice, see, for example, \citep{RevModPhys.80.517,RevModPhys.82.277,LAFLORENCIE20161,https://doi.org/10.1002/andp.202200064}.
Through the reduced density matrix and its Schmidt decomposition one
accesses to ``details'' of the entanglement - the eigenfunctions
and the spectrum, which allow, for example, to detect and characterize
topological order \citep{PhysRevLett.101.010504}.

The entanglement (or modular) Hamiltonian $H_{E}$, being a logarithm
of the RDM, carries the same information but is often preferable because
it appears to be ``more local'' than the reduced density matrix
itself, and allows alternative ways of calculation. Moreover, in some
cases, the entanglement Hamiltonians are known exactly. The most famous
and historically first example is given by the Bisognano-Wichmann
theorem \citep{10.1063/1.522605,10.1063/1.522898} which provides
an explicit and simple answer for a half-space entanglement Hamiltonian
in the ground state of an arbitrary Lorentz-invariant field theory
as an operator with a local density which is the Hamiltonian (energy)
density operator multiplied by the distance from the boundary. For
an arbitrary special region in a $d$-dimensional space-time, $H_{E}$
is in general non-local. The local terms in $H_{E}$ and their connection
to the Hamiltonian density were studied in \citep{PhysRevD.95.065005}
with the result that for $d\geq3$ local operators in $H_{E}$ are
in general not related to the energy density operator. Another examples
of the entanglement Hamiltonian for a half-space are provided by exactly
solvable two-dimensional lattice models \citep{RePEc:eee:phsmap:v:154:y:1988:i:1:p:1-20,Truong1989,https://doi.org/10.1002/andp.19995110203,Peschel_2004,Eisler_2020}
where the locality of the entanglement Hamiltonian is the same as
of the system's Hamiltonian. For in $1+1$ dimensional field theories
and for 1d lattice models, several results for $H_{E}$ are also available
for an interval and for a set of disjoint intervals. These include
exact results for conformal field theories \citep{Hislop1982,Casini2011,Cardy_2016},
free chiral scalar field \citep{PhysRevD.98.125008}, and massless
Dirac field \citep{PhysRevD.95.065005,Casini2009}. In the latter
case, also correction to $H_{E}$ in the small mass limit were also
found. For a lattice version of the above fields, as well as for some
other lattice systems, the structure of $H_{E}$ was discussed in
\citep{IngoPeschel_1999,PhysRevA.70.052329,Peschel_2009,Eisler_2018,DiGiulio_2020,Eisler_2020,Eisler_2022}.
In the above mentioned cases of $1+1$ dimensional massless field
theories, the progress in obtaining results for $H_{E}$ is heavily
based on underlying analytic structure of fields as functions of complex
space-time variables. A finite mass of the field, however, breaks
down this analytic structure and makes theoretical considerations
of the continuum case more complicated \citep{PhysRevD.96.105019}.

In this paper, an exact solution for the ground state reduced density
matrix on a finite interval for a free real scalar field in $1+1$
dimensions is presented and then used to discuss the corresponding
reduced density matrix. The solution covers an entire interval of
field's mass, including the massless case where it allows to construct
an unambiguous scheme of the Williamson diagonalization of the correlation
functions and reproduces the expression for $H_{E}$ known from conformal
field theory. Additionally, this scheme makes possible an expansion
for $E_{H}$ in the small mass case to obtain (nonlocal) corrections
to zero-mass result. In the opposite limit of a large mass, based
on the inverse mass expansion, it is shown that the effective temperature
ansatz for the local terms in the entanglement Hamiltonian is consistent
with the inverse effective temperature being linear in the distance
from the nearby edge.

The paper is organized as follows. In Section \ref{sec:Exact-solution-for RDM}
we present an exact solution for the reduced density matrix for an
interval and check the solution by calculating the correlation functions
for the well-known case of a massless field. Based on this solution,
in Section \ref{sec:Local-ansatz-for EH} arguments are provided for
the local entanglement Hamiltonian density is both massless and infinite
mass cases. Section \ref{sec:Diagonalization-of-the RDM} contains
the description of the Williamson diagonalization, which is performed
in Section \ref{sec:Diagonalization-for-M=00003D0} for the massless
case. Based on the results of the diagonalization, the leading order
corrections to the entanglement Hamiltonian are found in Section \ref{sec:limit Mto0.}
for the small mass case. Finally, conclusion and discussion are presented
in Section \ref{sec:Discussion-and-Conclusions}.

\section{Exact solution for the reduced density matrix\label{sec:Exact-solution-for RDM}}

Let us consider a one-dimensional real scalar field of mass $M_{\phi}$
described by the Hamiltonian
\begin{equation}
\hat{H}=\frac{1}{2}\int dx\{\hat{\pi}_{x}^{2}+(\partial_{x}\hat{\phi}_{x})^{2}+M_{\phi}^{2}\hat{\phi}_{x}^{2}\},\label{eq:H}
\end{equation}
where the field operator $\hat{\phi}_{x}$ and its conjugate momentum
operator $\hat{\pi}_{x}$ satisfy the canonical commutation relation
$[\hat{\pi}_{x},\hat{\phi}_{x'}]=-i\delta(x-x')$. Our goal is to
find the ground state reduced density matrix (modular Hamiltonian
in the field theory language) for an interval $\mathrm{I}_{l}$ of
length $2l$, $-l\leq x\leq l$. 

We start with the well-known expression for the matrix element of
the ground state reduced density matrix in terms of the functional
integral
\begin{equation}
\bra{\phi_{1}}\hat{\rho}\ket{\phi_{2}}\sim\int D\phi\exp[-S\{\phi\}],\label{eq:DM functional integral}
\end{equation}
where $\phi_{1}(x)$ and $\phi_{2}(x)$are field configurations on
$\mathrm{I}_{l}$, which satisfy the conditions $\phi_{1}(-l)=\phi_{2}(-l)$
and $\phi_{1}(l)=\phi_{2}(l)$, and
\begin{equation}
S\{\phi\}=\frac{1}{2}\int d\tau dx[(\partial_{\tau}\phi)^{2}+(\partial_{x}\phi)^{2}+M_{\phi}^{2}\phi^{2}]\label{eq:action}
\end{equation}
is the Euclidean action. The integration in (\ref{eq:DM functional integral})
is performed over fields $\phi(\tau,x)$ which are (a) defined on
the two-dimensional plane ($\tau$, $x$) with a cut \{$\tau=0,$$-l\leq x\leq l$\},
(b) decay at infinity, and (c) fulfill the boundary conditions $\phi(\tau=+0,x)=\phi_{1}(x)$
on the upper edge of the cut and $\phi(\tau=-0,x)=\phi_{2}(x)$. In
what follows, $l$ will be used as a unit of length meaning that the
cut corresponds to $-1\leq x\leq1$, and $M=M_{\phi}l$ is the dimensional
mass. Note that the normalization coefficient of the reduced density
matrix, being field-independent, is not important for our purposes
and, therefore, is not written down explicitly.

The Gaussian integral (\ref{eq:DM functional integral}) can be performed
with the result
\begin{equation}
\bra{\phi_{1}}\hat{\rho}\ket{\phi_{2}}\sim\exp[-S\{\phi_{cl}\}],\label{eq:DM intermediate}
\end{equation}
where the saddle point (classical) solution $\phi_{cl}(\tau,x)$ satisfy
the equation
\begin{equation}
[\partial_{\tau}^{2}+\partial_{x}^{2}-M^{2}]\phi_{cl}=0,\label{eq:LE original}
\end{equation}
decays at infinity, and fulfills the boundary conditions
\begin{equation}
\left\{ \begin{array}{cc}
\phi_{cl}(\tau=+0,x)= & \phi_{1}(x)\\
\phi_{cl}(\tau=-0,x)= & \phi_{2}(x)
\end{array}\right.\label{eq:BC}
\end{equation}
for $-1\leq x\leq1$. After using the Gaussian theorem and Eq.~(\ref{eq:LE original}),
the expression for $S\{\phi_{cl}\}$ can be written as

\begin{widetext}
\begin{equation}
S\{\phi_{cl}\}=-\frac{1}{2}\int_{-1}^{1}dx[\phi_{1}(x)\partial_{\tau}\phi_{cl}(\tau,x)|_{\tau=+0}-\phi_{2}(x)\partial_{\tau}\phi_{cl}(\tau,x)|_{\tau=-0}],\label{eq:S0}
\end{equation}
\end{widetext}where we use the fact that the outer normal derivative
to the area of integration corresponds to $-\partial_{\tau}$ on the
upper edge and to $\partial_{\tau}$ on the lower edge of the cut.

\subsection{Saddle point solution $\phi_{cl}(\tau,x)$}

To find $\phi_{cl}(\tau,x)$, it is convenient to introduce new variable
$u\geq0$ (``radial'') and $v\in[0,2\pi]$ (``angular'') according
to 
\[
x=\cosh u\cos v,\:\tau=\sinh u\sin v,
\]
or in the complex notations $z=x+i\tau=\cosh w$ with $w=u+iv$. In
the new variables, the cut corresponds to $u=0$ with $v\in[0,\pi]$
on the its upper edge and $v\in[\pi,2\pi]$ on the lower one. Equation
(\ref{eq:LE original}) transforms into
\begin{equation}
[\partial_{u}^{2}+\partial_{v}^{2}-\frac{1}{2}M^{2}(\cosh2u-\cos2v)]\phi_{cl}(u,v)=0\label{eq:LEnew}
\end{equation}
and the boundary conditions (\ref{eq:BC}) into
\begin{equation}
\left\{ \begin{array}{cc}
\phi_{cl}(u=0,v)= & \phi_{1}(v),\:v\in[0,\pi]\\
\phi_{cl}(u=0,v)= & \phi_{2}(v),\:v\in[\pi,2\pi]
\end{array}\right.,\label{eq:BCnew}
\end{equation}
where the same notation is kept for the boundary field values in new
angular variable, $\phi_{1(2)}(v)=\phi_{1(2)}(x=\cos v)$. Note that
the points $v=0$ and $v=2\pi$ are mapped onto the same point $x=1,$and
therefore, $\phi_{cl}(u=0,0)=\phi_{cl}(u=0,2\pi)$ such that the boundary
condition can be periodically extend to $\nu\in(-\infty,\infty)$. 

As it is well-known, equation (\ref{eq:LEnew}) allows separation
of variables, $\phi_{cl}(u,v)=f(u)g(v)$, with the following equations
for the functions $f(u)$ and $g(v)$
\begin{align}
[\partial_{v}^{2}+(\lambda+\frac{1}{2}M^{2}\cos2v)]g(v)= & 0,\label{eq:g}\\{}
[\partial_{u}^{2}-(\lambda+\frac{1}{2}M^{2}\cosh2u)]f(u)= & 0,\label{eq:f}
\end{align}
where $\lambda$ is the separation parameter. The above equations
coincide with the standard equations for the Mathieu functions {[}for
the function $g(v)${]} and modified Mathieu functions {[}for $f(u)]$
in the notations of \citep{mclachlan1947theory},
\begin{align}
[\partial_{v}^{2}+(\lambda-2q\cos2v)]g(v)= & 0,\label{eq:gstandard}\\{}
[\partial_{u}^{2}-(\lambda-2q\cosh2u)]f(u)= & 0,\label{eq:fstandard}
\end{align}
if one substitute $q$ with $-q=-(M/2)^{2}$. In the considered problem,
$g(v)$ has to satisfy the condition $g(v+2\pi)=g(v)$ and $f(u)$
has to decay for $u\to\infty$. Such solutions exist only for specific
values of the parameter $\lambda$ (eigenvalues $a_{m}$or $b_{m}$)
and are of the following forms (see \citep{mclachlan1947theory} and
Appendix~\ref{App:Mathieu functions}):
\[
\mathrm{Fek}_{m}(u,-q)\mathrm{ce}_{m}(v,-q)
\]
with eigenvalues $\lambda=a_{2n}$ for $m=2n$ and $\lambda=b_{2n+1}$
for and $m=2n+1$, and 
\[
\mathrm{Gek}_{m}(u,-q)\mathrm{se}_{m}(v,-q)
\]
with eigenvalues $\lambda=a_{2n+1}$ for $m=2n+1$ and $\lambda=b_{2n+2}$
for $m=2n+2$. Here $n$ is a non-negative integer, $n=0,1,\ldots$.
As functions of $v$, $\mathrm{ce}_{m}(v,-q)$ are even, while $\mathrm{se}_{m}(v,-q)$
are odd, both are normalized to $\pi$ on the interval $0\leq v\leq2\pi$
and form a complete set on this interval. For even $m$ they are $\pi$-periodic
and $2\pi$-periodic for odd $m$. 

The general solution of Eq. (\ref{eq:LEnew}) can therefore be written
as\begin{widetext}
\begin{equation}
\phi_{cl}(u,v)=\sum_{m}\{c_{m}\mathrm{Fek}_{m}(u,-q)\mathrm{ce}_{m}(v,-q)+d_{m}\mathrm{Gek}_{m}(u,-q)\mathrm{se}_{m}(v,-q)\},\label{eq:general solution}
\end{equation}
\end{widetext}where the constants $c_{m}$and $d_{m}$ are determined
from the boundary conditions (\ref{eq:BCnew}),
\[
c_{m}=\frac{1}{\pi\mathrm{Fek}_{m}(0,-q)}\int_{0}^{2\pi}\phi_{cl}(u=0,v)\mathrm{ce}_{m}(v,-q)dv
\]
and
\[
d_{m}=\frac{1}{\pi\mathrm{Gek}_{m}(0,-q)}\int_{0}^{2\pi}\phi_{cl}(u=0,v)\mathrm{se}_{m}(v,-q)dv,
\]
where we have use the completeness of the set of functions $\{\mathrm{ce}_{m}(v,-q),\mathrm{se}_{m}(v,-q)\}$
on the interval $v\in[0,2\pi]$ and their normalization to $\pi$
(see Appendix~\ref{App:Mathieu functions}). After using symmetric
and periodic properties of the Mathieu functions $\mathrm{ce}_{m}(v,-q)$
and $\mathrm{se}_{m}(v,-q)$, the above expressions can be rewritten
as
\begin{equation}
c_{m}=\frac{1}{\pi\mathrm{Fek}_{m}(0,-q)}\int_{0}^{\pi}\phi_{+}(v)\mathrm{ce}_{m}(v,-q)dv\label{eq:cm}
\end{equation}
and
\begin{equation}
d_{m}=\frac{1}{\pi\mathrm{Gek}_{m}(0,-q)}\int_{0}^{\pi}\phi_{-}(v)\mathrm{se}_{m}(v,-q)dv\label{eq:dm}
\end{equation}
with $\phi_{\pm}(v)=\phi_{1}(v)\pm\phi_{2}(v)$. Note that, as a consequence
of the boundary conditions, $\phi_{-}(0)=\phi_{-}(\pi)=0$.

\subsection{Reduced density matrix}

To proceed with finding the matrix element of the reduced density
matrix (\ref{eq:DM intermediate}), we first rewrite its exponent
(\ref{eq:S0}) in variables $u$ and $v$:

\begin{widetext}
\[
S\{\phi_{cl}\}=-\frac{1}{2}[\int_{0}^{\pi}dv\phi_{1}(v)\partial_{u}\phi_{cl}(u,v)|_{u=0}+\int_{\pi}^{2\pi}dv\phi_{2}(x)\partial_{u}\phi_{cl}(u,v)|_{u=0}]=-\frac{1}{2}\int_{0}^{2\pi}dv\phi_{cl}(u=0,v)\partial_{u}\phi_{cl}(u,v)|_{u=0}.
\]
Then, after substituting the solution (\ref{eq:general solution})
and using orthogonality of the Mathieu functions, we obtain
\[
S\{\phi_{cl}\}=-\frac{1}{2}\pi\sum_{m}\{c_{m}^{2}\mathrm{Fek}_{m}(0,-q)\mathrm{Fek'}_{m}(0,-q)+d_{m}^{2}\mathrm{Gek}_{m}(0,-q)\mathrm{Gek'}_{m}(0,-q)\},
\]
where $\mathrm{Fek'}_{m}(0,-q)\equiv\partial_{u}\mathrm{Fek}_{m}(u,-q)|_{u=0}$
and $\mathrm{Gek'}_{m}(0,-q)\equiv\partial_{u}\mathrm{Gek}_{m}(u,-q)|_{u=0}$
are the derivatives calculated at $u=0$. Finally, using of expressions
(\ref{eq:cm}) and (\ref{eq:dm}) for $c_{m}$and $d_{m},$the result
for the matrix element can be written as
\begin{equation}
\bra{\phi_{1}}\hat{\rho}\ket{\phi_{2}}\sim\exp\{-\frac{1}{2}\intop_{0}^{\pi}dv_{1}\intop_{0}^{\pi}dv_{2}\{K_{c}(v_{1},v_{2})\phi_{+}(v_{1})\phi_{+}(v_{2})+K_{s}(v_{1},v_{2})\phi_{-}(v_{1})\phi_{-}(v_{2})\}\},\label{eq:RDM general}
\end{equation}
where
\begin{equation}
K_{c}(v_{1},v_{2})=-\frac{1}{\pi}\sum_{m=0}\frac{\mathrm{Fek'}_{m}(0,-q)}{\mathrm{Fek}_{m}(0,-q)}\mathrm{ce}_{m}(v_{1},-q)\mathrm{ce}_{m}(v_{2},-q)\label{eq:Kc}
\end{equation}
and
\begin{equation}
K_{s}(v_{1},v_{2})=-\frac{1}{\pi}\sum_{m=1}\frac{\mathrm{Gek'}_{m}(0,-q)}{\mathrm{Gek}_{m}(0,-q)}\mathrm{se}_{m}(v_{1},-q)\mathrm{se}_{m}(v_{2},-q)\label{eq:Ks}
\end{equation}
\end{widetext}with $q=M^{2}/4$. Note that the kernels $K_{s}$ and
$K_{c}$ are related to the vacuum momentum-momentum $P$ and inverse
filed-field $Q^{-1}$ correlation functions, respectively. Namely,
\begin{equation}
K_{s}(x_{1},x_{2})=P(x_{1}-x_{2})\label{eq:Ks to P}
\end{equation}
and
\begin{equation}
K_{c}(x_{1},x_{2})=\frac{1}{4}(Q^{-1})(x_{1},x_{2}),\label{eq:Kc to inverse Q}
\end{equation}
in the original variables, see Appendix \ref{App:Reduced-density-matrix}.
In the following, both notations will be used depending on the context.

\subsection{Massless case\label{subsec:Massless-case}}

Let us test the above results in the known case of a massless field.
For $M=0$ one has $q=0$, and the Mathieu functions become trigonometric
functions, $\mathrm{ce}_{0}(v,-q)\to1/\sqrt{2}$, $\mathrm{ce}_{m}(v,-q)\to\cos mv$
and $\mathrm{se}_{m}(v,-q)\to\sin mv$. Functions $\mathrm{Fek}_{m}(u,-q)$
and $\mathrm{Gek}_{m}(u,-q)$ become decaying exponents, $\mathrm{Fek}_{m}(u,-q),\mathrm{Gek}_{m}(u,-q)\sim\exp(-mu)$.
We then obtain the following formal expressions for the kernels $K_{c}(v_{1},v_{2})$
and $K_{s}(v_{1},v_{2})$,
\begin{equation}
K_{c}^{(0)}(v_{1},v_{2})=\frac{1}{\pi}\sum_{m=1}m\cos mv_{1}\cos mv_{2}\label{eq:Kc0}
\end{equation}
and 
\begin{equation}
K_{s}^{(0)}(v_{1},v_{2})=\frac{1}{\pi}\sum_{m=1}m\sin mv_{1}\sin mv_{2}.\label{eq:Ks0}
\end{equation}
As usual, to perform summation in such series {[}and also in Eq.~(\ref{eq:Kc0-1})
below{]}, a convergency factor $\exp(-\alpha m)$ with $\alpha>0$
is introduced implying the limit $\alpha\to0$ in the end. One then
obtains
\begin{align}
K_{c}^{(0)}(v_{1},v_{2}) & =\frac{1}{2\pi}\frac{1-\cos v_{1}\cos v_{2}}{(\cos v_{1}-\cos v_{2})^{2}}\label{eq:Kc0final}\\
 & =\frac{1}{4\pi}\partial_{v_{1}}\partial_{v_{2}}\ln\frac{1-\cos(v_{1}+v_{2})}{1-\cos(v_{1}-v_{2})}\nonumber 
\end{align}
and
\begin{align}
K_{s}^{(0)}(v_{1},v_{2}) & =-\frac{1}{2\pi}\frac{\sin v_{1}\sin v_{2}}{(\cos v_{1}-\cos v_{2})^{2}}\label{eq:Ks0final}\\
 & =-\frac{1}{2\pi}\partial_{v_{1}}\partial_{v_{2}}\ln[2|\cos v_{1}-\cos v_{2}|].\nonumber 
\end{align}
As it was mentioned before, see Eq.~(\ref{eq:Ks to P}), the kernel
$K_{s}^{(0)}(v_{1},v_{2})$ gives the momentum-momentum correlations
function $K_{s}^{(0)}(v_{1},v_{2})=\braket{\hat{\pi}_{v_{1}}\hat{\pi}_{v_{2}}}$
in the angular variables {[}with $\hat{\pi}_{v}=-i\delta/\delta\phi(v)${]}
and, being rewritten in the original coordinates $x_{i}=\cos v_{i}$,
takes the familiar form $P_{0}(x_{1},x_{2})=(-1/2\pi)(x_{1}-x_{2})^{2}$.
{[}The sinus functions in (\ref{eq:Ks0final}) take care of the momentum
operator transformation $\hat{\pi}_{x}=-i\delta/\delta\phi(x)\to(\sin v)^{-1}\hat{\pi}_{v}$
under the change of variables.{]} Note also that, after using the
expansion of the Mathieu functions for small $M$, see Appendix \ref{App:Mathieu functions},
one reproduces the small-mass expansion of the momentum-momentum correlation
function\begin{widetext}
\begin{align*}
K_{s}(x_{12}) & =-\frac{M^{2}}{2\pi}\frac{\mathrm{K}_{1}(M|x_{12}|)}{M|x_{12}|}=-\frac{M^{2}}{2\pi}\left[\mathrm{K}_{2}(M|x_{12}|)-\mathrm{K}_{0}(M|x_{12}|)\right]\\
 & \approx-\frac{1}{2\pi}\frac{1}{|x_{12}|^{2}}-\frac{M^{2}}{4\pi}\left[\ln\frac{M|x_{12}|}{2}-\gamma-\frac{1}{2}\right]+O(M^{4}|x_{12}|^{2}\ln[M|x_{12}]),
\end{align*}
\end{widetext}where $x_{12}=x_{1}-x_{2}$, $\mathrm{K}_{\nu}(z)$
is the modified Bessel function, and $\gamma$ the Euler constant.
This result, together with Eqs.~(\ref{eq:Ks}) and (\ref{eq:Ks to P}),
suggests the following expansion of the modified Bessel function in
terms of the Mathieu functions\begin{widetext}

\[
\mathrm{K}_{1}(M|\cos v_{1}-\cos v_{2}|)=\frac{2}{M}\frac{|\cos v_{1}-\cos v_{2}|}{\sin v_{1}\sin v_{2}}\sum_{m=1}\frac{\mathrm{Gek'}_{m}(0,-q)}{\mathrm{Gek}_{m}(0,-q)}\mathrm{se}_{m}(v_{1},-q)\mathrm{se}_{m}(v_{2},-q).
\]

\end{widetext}As it can be seen from Eqs. (\ref{eq:Kc0}) and (\ref{eq:Kc0final}),
the kernel $K_{c}^{(0)}(v_{1},v_{2})$ satisfy the condition
\[
\int_{0}^{\pi}K_{c}^{(0)}(v_{1},v_{2})dv_{1}=\int_{0}^{\pi}K_{c}^{(0)}(v_{1},v_{2})dv_{2}=0
\]
as a result of the presence of a zero mode $\psi_{0}(v)=\mathrm{const}$.
Therefore, the inverse of this kernel can be uniquely defined only
on the orthogonal to the zero mode subspace. According to Eq.~(\ref{eq:Kc0}),
the formal inverse of the kernel $K_{c}^{(0)}(v_{1},v_{2})$ on such
subspace is
\begin{align}
\left(K_{c}^{(0)-1}\right)(v_{1},v_{2}) & =\frac{4}{\pi}\sum_{m=1}\frac{1}{m}\cos mv_{1}\cos mv_{2}\label{eq:Kc0-1}\\
 & =-\frac{1}{\pi}\ln[4(\cos v_{1}-\cos v_{2})^{2}]\nonumber 
\end{align}
and corresponds up to a coefficient $4$, see Eq.~(\ref{eq:Kc to inverse Q}),
to the vacuum field-field correlation function $Q_{0}(x_{1},x_{2})=\braket{\hat{\phi}_{x_{1}}\hat{\phi}_{x_{2}}}=-(1/4\pi)\ln[4(x_{1}-x_{2})^{2}]$
in the original coordinates. Note that thus defined inverse $K_{c}^{(0)-1}$
satisfies the condition
\[
\int_{0}^{\pi}\left(K_{c}^{(0)-1}\right)(v_{1},v_{2})dv_{1}=\int_{0}^{\pi}\left(K_{c}^{(0)-1}\right)(v_{1},v_{2})dv_{2}=0
\]
which fixes an otherwise arbitrary constant coming from the zero mode
subspace. Being rewritten in the original units, Eq.~(\ref{eq:Kc0final})
gives the kernel $\left(Q_{0}^{-1}\right)(x_{1},x_{2})$ for the inverse
$\hat{Q}_{0}^{-1}$ of $\hat{Q}_{0}$, see Eq.~(\ref{eq:Kc to inverse Q}),
\[
\left(Q_{0}^{-1}\right)(x_{1},x_{2})=\frac{2}{\pi}\frac{1-x_{1}x_{2}}{(x_{1}-x_{2})^{2}}\frac{1}{\sqrt{(1-x_{1}^{2})(1-x_{2}^{2})}},
\]
where the square root in the denominator of the last factor is due
to the change of variables.

\section{Local terms in the entanglement Hamiltonian\label{sec:Local-ansatz-for EH}}

An entanglement (or modular in the field theory language) Hamiltonian
\textbf{(EH)}, $\hat{H}_{E}=-(2\pi)^{-1}\log\hat{\rho}$, is in general
a nonlocal operator with only few exceptions, for example, Lorenz
invariant theories in a half-space (Bisognano-Wichmann theorem \citep{10.1063/1.522605,10.1063/1.522898})
and conformal field theories in $1+1$ dimension on a finite interval
\citep{Cardy_2016}. In these cases, the entanglement Hamiltonian
density is the Hamiltonian density modulated by a position-dependent
function which is often called inverse local temperature \citep{PhysRevD.95.065005}.
This function is zero at the boundary and grows linearly with the
distance from the boundary for the half-space case or has an inverse-parabola
form in the case of an interval. Therefore, for the inverse local
temperature in our situation, one expects an inverted parabola for
$M=0$ and a triangular form in the opposite case $M\to\infty$ when
the edges are ``infinitely far'' from each other in units of the
system's correlation length (with possible deviations in the narrow
region $\sim M^{-1}$ in the middle of the interval for a finite but
large $M$). As it is shown below, the result (\ref{eq:RDM general})
for the matrix element of the reduced density matrix allows to verify
the locality and the form of the entanglement Hamiltonian in two limiting
cases: $M=0$ (reproducing the well-known result) and $M\to\infty$.
More precisely, it is demonstrated that the reduced density matrix
commutes with a local operator whose density is the Hamiltonian density
modulated by a function which has an inverse-parabola form for $M=0$
and a triangular one for $M\to\infty$.

In the considered case, the Hamiltonian density $\hat{h}_{x}=(1/2)[\hat{\pi}_{x}^{2}+(\partial_{x}\hat{\phi}_{x})^{2}+M^{2}\hat{\phi}_{x}^{2}]$,
rewritten in the angular variable $v$ ($x=\cos v$), takes the form
{[}with the corresponding integration measure $dv\sin(v)${]}
\[
\hat{h}_{v}=\frac{1}{2\sin^{2}v}[\hat{\pi}_{v}^{2}+(\partial_{v}\hat{\phi}_{v})^{2}+M^{2}\sin^{2}v\hat{\phi}_{v}^{2}].
\]
Let us consider a local operator of the form
\[
\hat{O}_{f}=\frac{1}{2}\intop_{0}^{\pi}dvf(v)[\hat{\pi}_{v}^{2}+(\partial_{v}\hat{\phi}_{v})^{2}+M^{2}\sin^{2}v\hat{\phi}_{v}^{2}]
\]
with some arbitrary function $f(v)$ satisfying the condition $f(v)=f(\pi-v)$
which follows from the symmetry of the problem, and $f(0)=0$. The
matrix element of the commutator of the operator $\hat{O}_{f}$ with
the reduced density matrix $\hat{\rho}$, see Eq.~(\ref{eq:RDM general}),
reads\begin{widetext}
\begin{align*}
\bra{\phi_{1}}[\hat{\rho},\hat{O}_{f}]\ket{\phi_{2}} & =\frac{1}{2}\intop_{0}^{\pi}dvf(v)\left\{ [\frac{\delta^{2}}{\delta\phi_{1}^{2}(v)}-\frac{\delta^{2}}{\delta\phi_{2}^{2}(v)}]+[\partial_{v}\phi_{2}(v)]^{2}-[\partial_{v}\phi_{1}(v)]^{2}+M^{2}\sin^{2}v[\phi_{2}^{2}(v)-\phi_{1}^{2}(v)]\right\} \rho_{\phi_{1}\phi_{2}}\\
 & =\frac{1}{2}\intop_{0}^{\pi}dvf(v)\left[4\frac{\delta^{2}}{\delta\phi_{+}(v)\delta\phi_{-}(v)}-\partial_{v}\phi_{+}(v)\partial_{v}\phi_{-}(v)-M^{2}\sin^{2}v\phi_{+}(v)\phi_{-}(v)\right]\rho_{\phi_{1}\phi_{2}}\\
 & =\frac{1}{2}\intop_{0}^{\pi}dvf(v)\left[4(\hat{K}_{c}\phi_{+})(v)(\hat{K}_{s}\phi_{-})(v)-\partial_{v}\phi_{+}(v)\partial_{v}\phi_{-}(v)-M^{2}\sin^{2}v\phi_{+}(v)\phi_{-}(v)\right]\rho_{\phi_{1}\phi_{2}},
\end{align*}
where $(\hat{K}_{c/s}\phi_{\pm})(v)=\intop_{0}^{\pi}dv'K_{c/s}(v,v')\phi_{\pm}(v')$
and $\rho_{\phi_{1}\phi_{2}}=\bra{\phi_{1}}\hat{\rho}\ket{\phi_{2}}$.
Using the fact that the sets \{$\mathrm{ce}_{m}(z,-q)$\} and \{$\mathrm{se}_{m}(z,-q)$\}
are both complete on the interval $0<v<\pi$, one can take $\phi_{+}(v)=\mathrm{ce}_{m_{1}}(v,-q)$
and $\phi_{-}(v)=\mathrm{se}_{m_{2}}(v,-q)$. Then the integrals over
$v_{1}$ and $v_{2}$ in the first term of the last line can be performed
with the result
\begin{align}
\bra{\phi_{1}}[\hat{\rho},\hat{O}_{f}]\ket{\phi_{2}} & =\frac{1}{2}\intop_{0}^{\pi}dvf(v)\Biggl\{\biggl[\frac{\mathrm{Fek'}_{m_{1}}(0)}{\mathrm{Fek}_{m_{1}}(0)}\frac{\mathrm{Gek'}_{m_{2}}(0)}{\mathrm{Fek}_{m_{2}}(0)}-M^{2}\sin^{2}v\biggr]\mathrm{ce}_{m_{1}}(v)\mathrm{se}_{m_{2}}(v)\label{eq:commutator matrix element}\\
 & -\partial_{v}\mathrm{ce}_{m_{1}}(v)\partial_{v}\mathrm{se}_{m_{2}}(v)\Biggr\}\rho_{\phi_{1}\phi_{2}},\nonumber 
\end{align}
\end{widetext}where the argument $-q$ in the Mathieu functions is
omitted. From the condition $f(v)=f(\pi-u)$ and the properties of
the Mathieu functions (see Appendix~\ref{App:Mathieu functions})
it follows that the integral over $v$ is nonzero only if $m_{1}$and
$m_{2}$ have opposite parities, i.e., when either $m_{1}=2n_{1}$
and $m_{2}=2n_{2}+1$ (eigenvalues $a_{m_{1}}$and $a_{m_{2}}$) or
$m_{1}=2n_{1}+1$ and $m_{2}=2n_{2}+2$ (eigenvalues $b_{m_{1}}$and
$b_{m_{2}}$), where $n_{1},n_{2}\geq0$. With this in mind, we can
now analyze the cases $M=0$ and $M\to\infty$.

For $M=0$, with the expressions for Mathieu functions for this case
(see Appendix~\ref{App:Mathieu functions}), the curly bracket in
Eq.~(\ref{eq:commutator matrix element}) becomes $\{\ldots\}=m_{1}m_{2}\sin[(m_{1}+m_{2})v]$
and, when choosing $f(v)=\sin v$, the integral over $\nu$ gives
zero for all involved values of $m_{1}$ and $m_{2}$. It is easy
to see that for any other function $f(\nu)$ satisfying the above
mentioned conditions, some of the matrix elements will be nonzero.
As a result, the operator $\hat{O}_{f}$ which commutes with $\hat{\rho}$
has the form
\begin{align*}
\hat{O}_{f} & =\frac{1}{2}\intop_{0}^{\pi}dv\sin v[\hat{\pi}_{v}^{2}+(\partial_{v}\hat{\phi}_{v})^{2}]\\
 & =\frac{1}{2}\intop_{-1}^{1}dx(1-x^{2})[\hat{\pi}_{x}^{2}+(\partial_{x}\hat{\phi}_{x})^{2}]
\end{align*}
and just by a factor differs from the well-known entanglement Hamiltonian
in the massless case.

For the analysis of the $M\to\infty$ case, it is convenient to rewrite
Eq.~(\ref{eq:commutator matrix element}) using the following identity
\begin{align*}
 & \intop_{0}^{\pi}dvf(v)[\partial_{v}\mathrm{ce}_{m_{1}}(v)\partial_{v}\mathrm{se}_{m_{2}}(v)\\
 & \qquad\qquad\qquad\qquad\qquad+M^{2}\sin^{2}v\mathrm{ce}_{m_{1}}(v)se_{m_{2}}(v)]\\
 & =\intop_{0}^{\pi/2}dvf(v)[\frac{f''(v)}{f(v)}+\lambda_{m_{1}}+\lambda_{m_{2}}+4q]\mathrm{ce}_{m_{1}}(v)se_{m_{2}}(v),
\end{align*}
which follows from the equation on the Mathieu functions and is valid
for both cases $m_{1}=2n_{1}$, $m_{2}=2n_{2}+1$ (eigenvalues $a_{m_{1}}$and
$a_{m_{2}}$) and $m_{1}=2n_{1}+1$, $m_{2}=2n_{2}+2$ (eigenvalues
$b_{m_{1}}$and $b_{m_{2}}$). Here $\lambda_{m}$ are the corresponding
eigenvalues, $f''(v)$ is the second derivative of $f(v)$, and we
use the symmetry of the integrand under $v\to\pi-v$. With this result,
Eq.~(\ref{eq:commutator matrix element}) reads\begin{widetext}
\begin{equation}
\bra{\phi_{1}}[\hat{\rho},\hat{O}_{f}]\ket{\phi_{2}}=\frac{1}{2}\intop_{0}^{\pi}dvf(v)\mathrm{ce}_{m_{1}}(v)se_{m_{2}}(v)\left\{ \frac{\mathrm{Fek'}_{m_{1}}(0)}{\mathrm{Fek}_{m_{1}}(0)}\frac{\mathrm{Gek'}_{m_{2}}(0)}{\mathrm{Fek}_{m_{2}}(0)}-\frac{1}{2}\left[\frac{f''(v)}{f(v)}+\lambda_{m_{1}}+\lambda_{m_{2}}+4q\right]\right\} \rho_{\phi_{1}\phi_{2}}.\label{eq:commutator matrix element 1}
\end{equation}

In the limit $M\to\infty$, one has for both cases $m_{1}=2n_{1}$,
$m_{2}=2n_{2}+1$ and $m_{1}=2n_{1}+1$, $m_{2}=2n_{2}+2$ {[}see
Eqs.~(\ref{eq:logderFekGek}), (\ref{eq:am large q}) and (\ref{eq:gamma1})
in Appendix~\ref{App:Mathieu functions}{]}
\begin{equation}
\frac{\mathrm{Fek'}_{m_{1}}(0)}{\mathrm{Fek}_{m_{1}}(0)}\frac{\mathrm{Gek'}_{m_{2}}(0)}{\mathrm{Fek}_{m_{2}}(0)}=8q^{1/2}\frac{\Gamma(n_{1}+1)}{\Gamma(n_{1}+1/2)}\frac{\Gamma(n_{2}+3/2)}{\Gamma(n_{2}+1)}\left[1-\frac{n_{1}+n_{2}+1}{8q^{1/2}}+O(q^{-1})\right]\label{eq:product log der.}
\end{equation}
and
\begin{equation}
a/b_{m_{1}}+a/b_{m_{2}}+4q=8q^{1/2}[n_{1}+n_{2}+1+\frac{\gamma_{(1)2n_{1}}+\overline{\gamma}_{(1)2n_{2}+1}}{8q^{1/2}}+O(q^{-1})].\label{eq:a+a+4q}
\end{equation}
\end{widetext}Let us consider first the evaluation of the expression
(\ref{eq:commutator matrix element 1}) in the leading order in $q^{1/4}\sim M^{1/2}$.
In doing this we take the leading terms in the Mathieu functions expansions
in the variable $\alpha=2q^{1/4}\sin v$ through the parabolic cylinder
functions $\mathrm{D}_{p}(\alpha)$, see Appendix~\ref{App:Mathieu functions}.
Due to already mentioned symmetry of the problem, one can limit the
integration over $v$ to the interval $[0,\pi/2]$ removing also the
coefficient $1/2$ in front of the integral. Then one has
\begin{align*}
\intop_{0}^{\pi/2}dv\ldots & =\intop_{0}^{\pi/2}\frac{d(\sin v)}{\cos v}\ldots\\
 & =\frac{1}{2q^{1/4}}\intop_{0}^{2q^{1/4}}\frac{d\alpha}{\sqrt{1-\alpha^{2}/4q^{1/2}}}\ldots\\
 & \rightarrow\frac{1}{2q^{1/4}}\intop_{0}^{\infty}d\alpha[1+\frac{\alpha^{2}}{8q^{1/2}}+O(q^{-1})]\ldots,
\end{align*}
where we take into account that the relevant values of $\alpha$ are
of order $1$. The latter means that corresponding values of $v$
are much smaller then one, $v\ll1$. As a result, only the first (linear)
term in the expansion of the function $f(v)$, namely $f(v)\approx f'(0)v$,
contributes, and one can also neglect the term $f''(v)/f(v)=O(1)$
in the square brackets. The resulting integral over $v$ then reads
{[}see Eqs.~(\ref{eq:ce2n large q}) and (\ref{eq:se2n+1 large q})
in Appendix \ref{App:Mathieu functions}{]}\begin{widetext}
\begin{align*}
\intop_{0}^{\pi/2}dvf(v)\mathrm{ce}_{m_{1}}(v,-q)se_{m_{2}}(v,-q) & \to\frac{f'(0)}{4q^{1/2}}(-1)^{n_{1}+n_{2}}C_{2n_{1}}C_{2n_{2}+1}\intop_{0}^{\infty}d\alpha\alpha\mathrm{D}_{2n_{1}}(\alpha)\mathrm{D}_{2n_{2}+1}(\alpha)\\
 & =f'(0)\frac{\pi}{8q^{1/4}}[\sqrt{2n_{1}+1}\delta_{n_{1},n_{2}}-\sqrt{2n_{2}+2}\delta_{n_{1},n_{2}+1}],
\end{align*}
\end{widetext}where Eqs.~(\ref{eq:zDp}) and (\ref{eq:normDp})
together with the orthogonality of the parabolic cylinder functions
$\mathrm{D}_{p}(\alpha)$ when their orders have the same parity are
used. The leading term in the curly bracket in (\ref{eq:commutator matrix element 1})
is
\[
8q^{1/2}\left[\frac{\Gamma(n_{1}+1)}{\Gamma(n_{1}+1/2)}\frac{\Gamma(n_{2}+3/2)}{\Gamma(n_{2}+1)}-\frac{1}{2}(n_{1}+n_{2}+1)\right],
\]
so that in the leading order one finally obtains\begin{widetext}
\begin{align*}
\bra{\phi_{1}}[\hat{\rho},\hat{O}_{f}]\ket{\phi_{2}} & =\rho_{\phi_{1}\phi_{2}}\pi f'(0)q^{1/4}\biggl[\frac{\Gamma(n_{1}+1)}{\Gamma(n_{1}+1/2)}\frac{\Gamma(n_{2}+3/2)}{\Gamma(n_{2}+1)}-\frac{1}{2}(n_{1}+n_{2}+1)\biggr]\\
 & \times[\sqrt{2n_{1}+1}\delta_{n_{1},n_{2}}-\sqrt{2n_{2}+2}\delta_{n_{1},n_{2}+1}]=0.
\end{align*}
\end{widetext}As a result, to order $O(q^{1/4})\sim O(M^{1/2})$,
any operator $\hat{O}_{f}$ with $f(v)$ which is linear in $\nu$
for small $\nu$ (and in $\pi-\nu$ for $\nu\approx\pi$) commutes
with the reduced density matrix $\hat{\rho}$. In the $x$-variable,
this corresponds to functions linear in $1-x$ for $x\approx1$ and
(symmetrically) in $1+x$ for $x\approx-1$.

Further constraints on the function $f$ comes from the calculate
of the matrix element (\ref{eq:commutator matrix element 1}) to the
next order ($\sim q^{-1/4}$). In particular, this allows to distinguish
between two potential forms of $f$ - parabolic $\sim(1-x^{2})$ with
$f_{1}(v)=\sin v$ and triangular $\sim1-|x|$ with $f_{2}(v)=2(1-|\cos v|)/\sin v$
(here the factor $2$ is introduced for the later convenience), see
Ref.~\citep{PhysRevD.95.065005} and \citep{Eisler_2020} for the
continuum and lattice cases, respectively.

In doing this, one has to expand the curly bracket in (\ref{eq:commutator matrix element 1})
up to $O(1)$ terms and the integrand over $v$ up to $O(q^{-1/2})$
ones. Keeping in mind that 
\begin{equation}
\frac{f''(v)}{f(v)}=\beta+O(q^{-1/2}),\label{eq:beta}
\end{equation}
with $\beta=-1$ for $f_{1}(v)$ and $\beta=1/2$ for $f_{2}(v)$
(as before, we consider$v\in[0,\pi/2]$), the curly bracket in (\ref{eq:commutator matrix element 1})
can be written as {[}see Eqs.~(\ref{eq:a+a+4q}), (\ref{eq:product log der.}),
and (\ref{eq:beta}){]}
\begin{equation}
\left\{ \ldots\right\} =8q^{1/2}\biggl[\left\{ \ldots\right\} _{(0)}+\frac{1}{4q^{1/2}}\left\{ \ldots\right\} _{(1)}+O(q^{-1})\biggr],\label{eq:curly braket expansion}
\end{equation}
where
\begin{equation}
\left\{ \ldots\right\} _{(0)}=\frac{\Gamma(n_{1}+1)}{\Gamma(n_{1}+1/2)}\frac{\Gamma(n_{2}+3/2)}{\Gamma(n_{2}+1)}-\frac{1}{2}(n_{1}+n_{2}+1)\label{eq:curly braket 0}
\end{equation}
and\begin{widetext}
\begin{equation}
\left\{ \ldots\right\} _{(1)}=-\frac{1}{2}\left[\frac{\Gamma(n_{1}+1)}{\Gamma(n_{1}+1/2)}\frac{\Gamma(n_{2}+3/2)}{\Gamma(n_{2}+1)}(n_{1}+n_{2}+1)+\frac{1}{2}(\gamma_{(1)2n_{1}}+\overline{\gamma}_{(1)2n_{2}+1}+\beta)\right].\label{curly braket 1}
\end{equation}
The integral in (\ref{eq:commutator matrix element 1}) can be expanded
as 
\begin{equation}
\intop_{0}^{\pi/2}dvf(v)\mathrm{ce}_{m_{1}}(v,-q)se_{m_{2}}(v,-q)=\frac{1}{2q^{1/4}}\intop_{0}^{\infty}d\alpha\alpha[1+\delta\frac{\alpha^{2}}{8q^{1/2}}+O(q^{-1})]\mathrm{ce}_{m_{1}}(v,-q)se_{m_{2}}(v,-q),\label{integral expansion}
\end{equation}
where $\delta=1$ for $f_{1}(v)$ and $\delta=3/2$ for $f_{2}(v)$.
Finally, the expansion of the product of the Mathieu functions in
the above expression reads
\begin{align}
\mathrm{ce}_{m_{1}}(v,-q)se_{m_{2}}(v,-q) & =(-1)^{n_{1}+n_{2}}C_{2n_{1}}C_{2n_{2}+1}\Biggl\{\mathrm{D}_{2n_{1}}(\alpha)\mathrm{D}_{2n_{2}+1}(\alpha)\label{eq:cese expansion}\\
 & +\frac{1}{4q^{1/2}}\left[\mathrm{D}_{2n_{1}}(\alpha)\overline{y}_{(1)n_{2}}(\alpha)+\mathrm{D}_{2n_{2}+1}(\alpha)y_{(1)n_{1}}(\alpha)\right]+O(q^{-1})\Biggr\}
\end{align}
for both cases, see Appendix \ref{App:Mathieu functions}. For the
integral over $\alpha$ in (\ref{eq:commutator matrix element 1})
one then obtains

\[
(-1)^{n_{1}+n_{2}}C_{2n_{1}}C_{2n_{2}+1}\intop_{0}^{\infty}d\alpha\alpha\Biggl[\left\{ \ldots\right\} _{(0)}\mathrm{D}_{2n_{1}}\mathrm{D}_{2n_{2}+1}
\]
\[
+\frac{1}{4q^{1/2}}\left(\left\{ \ldots\right\} _{(1)}\mathrm{D}_{2n_{1}}\mathrm{D}_{2n_{2}+1}+\left\{ \ldots\right\} _{(0)}[\frac{\delta}{2}\alpha^{2}\mathrm{D}_{2n_{1}}\mathrm{D}_{2n_{2}+1}+\mathrm{D}_{2n_{1}}\overline{y}_{(1)n_{2}}+\mathrm{D}_{2n_{2}+1}y_{(1)n_{1}}]\right)+O(q^{-1})\Biggr],
\]
\end{widetext}where the dependence on $\alpha$ in the functions
$\mathrm{D}_{2n_{1}}(\alpha)$, $y_{(1)n_{1}}(\alpha)$, and $\overline{y}_{(1)n_{2}}(\alpha)$
is omitted. The integration over $\alpha$ can be performed using
already mentioned orthogonality of the parabolic cylinder function
and the following formulae
\[
\alpha\mathrm{D}_{m}=\mathrm{D}_{m+1}+m\mathrm{D}_{m-1},
\]
\[
\alpha^{2}\mathrm{D}_{m}=(2m+1)\mathrm{D}_{m}+\mathrm{D}_{m+2}+m(m+1)\mathrm{D}_{m-2}.
\]
The calculations are straightforward but lengthy, and result in the
following: To the considered order $O(q^{-1/2})$, the integrals over
$\alpha$ are nonzero only for the following cases (a) $n_{1}=n_{2}$,
(b) $n_{1}=n_{2}+1$, (c) $n_{1}=n_{2}+2$, (d) $n_{1}=n_{2}+3$,
(e) $n_{2}=n_{1}+1$, and (f) $n_{2}=n_{1}+2$. The cases (d) and
(f) give zero contributions, the result of cases (c) and (e) is proportional
to $\delta/2-3/4$, and to $\beta-1/2$ in the cases (a) and (b).
Therefore, in the considered order, the matrix element of the commutator
(\ref{eq:commutator matrix element}) is zero for $f_{2}(v)=2(1-|\cos v|)/\sin v$
but not for $f_{1}(v)=\sin v$. Note that $\beta$ and $\delta$ can
be expressed through the derivatives of the function $f(x)$ as $\beta=[1-3f''(1)/f'(1)]/2$
and $\delta=3[1-f''(1)/f'(1)]/2$, showing that the both conditions
$\beta=1/2$ and $\delta=3/2$ are satisfied when $f''(1)=0$. This
excluding the parabolic form for $f(x)$ or, more generally, the (mass-independent)
second order term in the expansion of $f(x)$ near the edges. As a
consequence, the ``local temperature'' has more complicated analytic
structure then suggested in \citep{PhysRevD.95.065005}.

The above calculation show that, in the large $M$ limit, the operator
\begin{align*}
\hat{O}_{f} & =\intop_{0}^{\pi}dv\frac{1-|\cos v|}{\sin v}[\hat{\pi}_{v}^{2}+(\partial_{v}\hat{\phi}_{v})^{2}+M^{2}\sin^{2}v\hat{\phi}_{v}^{2}]\\
 & =\intop_{-1}^{1}dx(1-|x|)[\hat{\pi}_{x}^{2}+(\partial_{x}\hat{\phi}_{x})^{2}+M^{2}\hat{\phi}_{x}^{2}]
\end{align*}
commutes with the reduced density matrix up to terms $\sim q^{-1/4}\sim M^{-1/2}$.
This supports a triangular form of the entanglement Hamiltonian density
in the infinite mass limit, which is also consistent with the result
of numerical calculations in the equivalent lattice case \citep{Eisler_2020}.
Intuitively, this result is expected because, in the case when the
correlation length is much smaller than the size of the subsystem,
the properties in the region close to one edge are only exponentially
small affected by the presence of the other edge. Therefore, with
exponential accuracy they are identical to those in a half-infinite
system where the Bisognano-Wichmann theorem gives a linearly growing
entanglement Hamiltonian density. In is only in the central region
of the system, where the influences of both edges are comparable,
one expects the appearance of a curvature in $f(x)$.

\section{Diagonalization of the reduced density matrix and entanglement Hamiltonian\label{sec:Diagonalization-of-the RDM}}

A common procedure of finding the entanglement Hamiltonian for quadratic
theories uses their field-field and momentum-momentum correlation
functions, see, for example, \citep{IngoPeschel_2003,Peschel_2009}.
The result of the previous Sections for the reduced density matrix
suggest a different (but equivalent) procedure which based on the
momentum-momentum and inverse field-field correlation functions. The
procedure turns out to be more convenient for the analysis of the
small $M$ limit because both involved kernels $K_{c}$ and $K_{s}$
are well-behaved when $M\to0$, allowing therefore a small parameter
expansion. This is in contrast to the standard scheme where the field-field
correlation function contains a diverging constant $\sim\ln M$.

The underlying idea can easily be explained by considering a harmonic
oscillator of mass $\mu$ and frequency $\omega$ with the Hamiltonian
\[
\hat{H}=\frac{\hat{\pi}^{2}}{2\mu}+\frac{\mu\omega^{2}\hat{\phi}^{2}}{2},
\]
where $\hat{\pi}$ and $\hat{\phi}$ are the momentum and position
operators, respectively. The corresponding density matrix $\hat{\rho}\sim\exp\{-2\pi\hat{H}\}$
for the inverse temperature $\beta=2\pi$ has matrix elements\begin{widetext}
\begin{align*}
\bra{\phi_{1}}\hat{\rho}\ket{\phi_{2}} & =\sqrt{\frac{\mu\omega}{2\pi\sinh(2\pi\omega)}}\exp\left\{ -\frac{\mu\omega}{2\sinh(2\pi\omega)}\left[(\phi_{1}^{2}+\phi_{2}^{2})\cosh(2\pi\omega)-2\phi_{1}\phi_{2}\right]\right\} \\
 & =\sqrt{\frac{\mu\omega}{2\pi\sinh(2\pi\omega)}}\exp\left\{ -\frac{1}{4}\mu\omega[\tanh(\pi\omega)\phi_{+}^{2}+\coth(\pi\omega)\phi_{-}^{2}]\right\} ,
\end{align*}
\end{widetext}where $\phi_{\pm}=\phi_{1}\pm\phi_{2}$. After introducing
$\epsilon=2\pi\omega$, $\lambda=\coth(\epsilon/2)$, and defining
the mass as $\mu=4\pi/(\epsilon\lambda)$, one obtains 
\[
\exp\left\{ -\frac{1}{2}[\lambda^{-2}\phi_{+}^{2}+\phi_{-}^{2}]\right\} 
\]
for the exponent in the above expression and 
\[
2\pi\hat{H}=\frac{1}{2}\epsilon\left[\frac{\lambda}{2}\hat{\pi}^{2}+\frac{2}{\lambda}\hat{\phi}^{2}\right]
\]
for the Hamiltonian (multiplied by $2\pi$). The canonical symmetric
form for the harmonic oscillator Hamiltonian $2\pi\hat{H}=(\epsilon/2)\left[\hat{\pi}^{2}+\hat{\phi}^{2}\right]$
is achieved by rescaling the operators $\hat{\pi}\to\sqrt{2/\lambda}\hat{\pi}$
and $\hat{\phi}\to\sqrt{\lambda/2}\hat{\phi}$. The exponent in the
expression for the matrix element of the density matrix is then $\exp\left\{ -[\lambda^{-1}\phi_{+}^{2}+\lambda\phi_{-}^{2}]/2\right\} $. 

The physical difference between the two forms is in the chosen unit
of length: Distances in the canonical form are in units of the harmonic
oscillator length ($\sim\omega^{-1/2}$) which diverges for $\omega\to0$
($\lambda\to\infty$). This form therefore is not convenient for the
analysis of a small frequency case. In the initial form, on the other
hand, distances are measured in natural ($\omega$-independent) units,
allowing therefore a well-defined zero frequency limit with the free-particle
Hamiltonian $2\pi\hat{H}=\hat{\pi}^{2}/2$ and the density matrix
$\bra{\phi_{1}}\hat{\rho}\ket{\phi_{2}}\sim\exp\left\{ -\phi_{-}^{2}/2\right\} $.

Generalization to the case with many degrees of freedom is straightforward:
For the Hamiltonian
\begin{equation}
2\pi\hat{H}=\frac{1}{2}\sum_{s}\epsilon_{s}\left[\frac{\lambda_{s}}{2}\hat{\pi}_{s}^{2}+\frac{2}{\lambda_{s}}\hat{\phi}_{s}^{2}\right]\label{eq:EntHam}
\end{equation}
with $\epsilon_{s}\geq0$ and $\lambda_{s}=\coth(\epsilon_{s}/2)$,
the exponent in the matrix element of the density matrix is
\[
\bra{\{\phi_{s1}\}}\hat{\rho}\ket{\{\phi_{s2}\}}\sim\exp\left\{ -\frac{1}{2}\sum_{s}[\lambda_{s}^{-2}\phi_{s+}^{2}+\phi_{s-}^{2}]\right\} 
\]
where $\phi_{s\pm}=\phi_{s1}\pm\phi_{s2}$. As a result, finding the
Hamiltonian $\hat{H}$ is reduced to diagonalization of the density
matrix, i.e. to diagonalization of the two quadratic forms with the
kernels $K_{c}$ and $K_{s}$. 

In our case of the free scalar field on the interval, the matrix elements
of the reduced density matrix are given by (\ref{eq:RDM general}).
To make a comparison with the standard approaches easier, following
Appendix~\ref{App:Reduced-density-matrix}, we replace the notations
for the kernels of the quadratic forms from $K_{c}$ and $K_{s}$
to $Q$ and $P$, the field-field and momentum-momentum correlation
functions, respectively, $K_{c}\to(1/4)Q^{-1}$ and $K_{s}\to P$.
The matrix element (\ref{eq:RDM general}) then reads
\begin{equation}
\bra{\phi_{1}(x)}\hat{\rho}\ket{\phi_{2}(y)}\sim\exp[-(\frac{1}{2}\phi_{-}\hat{P}\phi_{-}+\frac{1}{8}\phi_{+}\hat{Q}^{-1}\phi_{+})]\label{eq:RDM with P and Q}
\end{equation}
with compact notations for the quadratic forms, for example, $\phi_{-}\hat{P}\phi_{-}=\iint dxdy\phi_{-}(x)P(x,y)\phi_{-}(y)$. 

A simultaneous diagonalization of the quadratic forms in the exponent
of Eq.~(\ref{eq:RDM with P and Q}) is equivalent to finding a biorthogonal
complete set of functions \{$\psi_{s}(x)$, $\varphi_{s}(x)$\},
\begin{align}
\int dx\psi_{s}^{*}(x)\varphi_{s'}(x) & =\delta_{ss'},\label{eq:complete set conditions}\\
\sum_{s}\psi_{s}(x)\varphi_{s}^{*}(y) & =\delta(x-y)\nonumber 
\end{align}
which satisfy the following set of equations
\begin{align}
\hat{P}\psi_{s} & =\varphi_{s},\label{eq: W1}\\
\hat{Q}^{-1}\psi_{s} & =\left[\frac{2}{\lambda_{s}}\right]^{2}\varphi_{s}.\label{eq:W2}
\end{align}
Using these functions, the spectral decompositions for the kernels
of $\hat{P}$ and $\hat{Q}^{-1}$ take the form
\begin{align}
P(x,y) & =\sum_{s}\varphi_{s}^{*}(x)\varphi_{s}(y),\label{eq:P}\\
(Q^{-1})(x,y) & =\sum_{s}\left[\frac{2}{\lambda_{s}}\right]^{2}\varphi_{s}^{*}(x)\varphi_{s}(y),\label{eq:Q-1}
\end{align}
and 
\begin{align}
Q(x,y) & =\sum_{s}(\lambda_{s}/2)^{2}\psi_{s}^{*}(x)\psi_{s}(y),\label{eq:Q}\\
(P^{-1})(x,y) & =\sum_{s}\psi_{s}^{*}(x)\psi_{s}(y).\label{eq:P-1}
\end{align}
 for the kernels of $\hat{Q}$ and $\hat{P}^{-1}$. The corresponding
entanglement Hamiltonian is then given by\begin{widetext}
\begin{equation}
2\pi\hat{H}_{E}=\frac{1}{2}\sum_{s}\epsilon_{s}\left[\frac{\lambda_{s}}{2}\hat{\pi}_{s}^{*}\hat{\pi}_{s}+\frac{2}{\lambda_{s}}\hat{\phi}_{s}^{*}\hat{\phi}_{s}\right]=2\pi\int dxdy\left[H_{E\pi}(x,y)\hat{\pi}_{x}\hat{\pi}_{y}+E_{E\phi}(x,y)\hat{\phi}_{x}\hat{\phi}_{y}\right]\label{eq:EntHamc}
\end{equation}
 \end{widetext}with $\lambda_{s}=\coth(\epsilon_{s}/2)$,
\begin{align*}
\hat{\pi}_{s} & =\int dx\psi_{s}(x)\hat{\pi}_{x},\\
\hat{\phi}_{s} & =\int dx\varphi_{s}(x)\hat{\phi}_{x},
\end{align*}
and
\begin{equation}
2\pi H_{E\pi}(x,y)=\frac{1}{2}\sum_{s}\epsilon_{s}\frac{\lambda_{s}}{2}\psi_{s}^{*}(x)\psi_{s}(y),\label{eq:HEpi general}
\end{equation}
\begin{equation}
2\pi H_{E\phi}(x,y)=\frac{1}{2}\sum_{s}\epsilon_{s}\frac{2}{\lambda_{s}}\varphi_{s}^{*}(x)\varphi_{s}(y).\label{eq:HEphi general}
\end{equation}
Note that, although $\lambda_{s}$ and $\epsilon_{s}$ are continuous,
we prefer to use discrete notations for intermediate calculations,
and restore continuous notations in the end. It is also convenient
to use complex-valued functions with $\psi_{s}^{*}(x)$ {[}$\varphi_{s}^{*}(x)${]}
being the complex conjugate of $\psi_{s}(x)$ {[}$\varphi_{s}(x)${]}.
We also mention that the zero mode in the$M=0$ case, which corresponds
to $s=0$, with $\lambda_{0}^{-1}=\epsilon_{0}=0$, is present in
(\ref{eq:P}) and (\ref{eq:P-1}), absent in (\ref{eq:Q-1}), and
has to be excluded from (\ref{eq:Q}) due to its divergence. 

It is easy to see the equivalence of the above procedure to the standard
one after rescaling the functions, $\psi_{s}(x)\to\sqrt{2/\lambda_{s}}\overline{\psi}_{s}(x)$
and $\varphi_{s}(x)\to\sqrt{\lambda_{s}/2}\overline{\varphi}_{s}(x)$,
(this does not change the biorthogonality) such that the Entanglement
Hamiltonian becomes
\[
2\pi\hat{H}_{E}=\frac{1}{2}\sum_{s}\epsilon_{s}\left[\hat{\overline{\pi}}_{s}^{*}\hat{\overline{\pi}}_{s}+\hat{\overline{\phi}}_{s}^{*}\hat{\overline{\phi}}_{s}\right],
\]
where the operators are constructed using functions $\overline{\psi}_{s}(x)$
and $\overline{\varphi}_{s}(x)$, and Eqs.~(\ref{eq: W1}) and (\ref{eq:W2})
transform into
\begin{align*}
\hat{P}\overline{\psi}_{s} & =\frac{\lambda_{s}}{2}\overline{\varphi}_{s},\\
\hat{Q}^{-1}\overline{\psi}_{s} & =\frac{2}{\lambda_{s}}\overline{\varphi}_{s}.
\end{align*}
Applying the operator with the kernel $\hat{Q}$ to the second equation,
we transform it into $\hat{Q}\overline{\varphi}_{s}=(\lambda_{s}/2)\overline{\psi}_{s}$
which, together with the first equation, form a standard set for finding
the entanglement Hamiltonian from $P$ and $Q$ correlation functions.
As it was already mentioned, the disadvantage of the standard scheme
is that the corresponding equation do not allow the limit $\lambda_{s}\to0$,
when the equations for the corresponding functions $\overline{\psi}_{s}$
and $\overline{\varphi}_{s}$ becomes meaningless. The scheme based
on Eqs. (\ref{eq: W1}) and (\ref{eq:W2}), on the other hand, does
not encounter any problem in this limit. We will see below that the
functions $\psi_{s}(x)$ and $\varphi_{s}(x)$ have continuous limit
when $M\to0,$such that the kernel $P(x,y)$ is non-singular and has
a well-defined inverse. In this limit, the kernel $(Q^{-1})(x,y)$
acquires a zero mode and, therefore, its inverse is uniquely defined
only on the subspace which is orthogonal to the zero mode. 

\section{Diagonalization for $M=0$.\label{sec:Diagonalization-for-M=00003D0}}

For the massless case $M=0$, it is convenient to modify the system
of equations (\ref{eq: W1}) and (\ref{eq:W2}) with the kernels $\hat{P}_{0}$
and $\hat{Q}_{0}^{-1}$ as follows: For $M=0$, the operator $\hat{Q}_{0}^{-1}$
has a zero mode (constant function) corresponding to $\lambda_{s}^{-1}=0$
with the consequence (in the angular variables $u$ and $v$ such
that $x=\cos u$ and $y=\cos v$)
\[
\intop_{0}^{\pi}du(Q_{0}^{-1})(u,v)=\intop_{0}^{\pi}dv(Q_{0}^{-1})(u,v)=0.
\]
\[
\intop_{0}^{\pi}du(Q_{0}^{-1})(u,v)=\intop_{0}^{\pi}dv(Q_{0}^{-1})(u,v)=0,
\]
see (\ref{eq:Kc0}) and (\ref{eq:Kc0final}). These imply that Eq.~(\ref{eq:W2})
can be written in the form
\begin{equation}
\hat{Q}_{0}^{-1}[\psi_{s}-\braket{\psi_{s}}]=\left[\frac{2}{\lambda_{s}}\right]^{2}\varphi_{s},\label{eq:W20modified}
\end{equation}
where
\[
\braket{\psi_{s}}=\frac{1}{\pi}\intop_{0}^{\pi}du\psi_{s}(u),
\]
such that the square bracket in (\ref{eq:W20modified}) has zero mean
value (orthogonal to the zero mode), and also that
\begin{equation}
\left[\frac{2}{\lambda_{s}}\right]^{2}\intop_{0}^{\pi}du\varphi_{s}(u)=0\label{eq:W20rhs}
\end{equation}
for all $s.$ Under this condition, after applying $\hat{Q}_{0}$
to Eq.~(\ref{eq:W20modified}), one obtains
\[
\hat{Q}_{0}\varphi_{s}=\left[\frac{\lambda_{s}}{2}\right]^{2}[\psi_{s}-\braket{\psi_{s}}].
\]

A more symmetric form of the final set of equation for the massless
case, which also allows a straightforward application of the approach
from Ref.~\citep{PhysRevD.98.125008}, can be obtained by using the
rescaled complete set of functions \{$\overline{\psi}_{s}(x)$, $\overline{\varphi}_{s}(x)$\}
(see end of Section \ref{sec:Diagonalization-of-the RDM}):

\begin{align}
\hat{P}_{0}\overline{\psi}_{s} & =\frac{\lambda_{s}}{2}\overline{\varphi}_{s},\label{eq:P0psi}\\
\hat{Q}_{0}\overline{\varphi}_{s} & =\frac{\lambda_{s}}{2}[\overline{\psi}_{s}-\braket{\overline{\psi}_{s}}],\label{eq:Q0phi}
\end{align}
where
\[
\braket{\overline{\psi}_{s}}=\frac{1}{\pi}\intop_{0}^{\pi}du\overline{\psi}_{s}(u).
\]
Eq.~() actually implies that the spectral decomposition of $\hat{Q}_{0}$
in the massless case has the form
\begin{equation}
Q_{0}(x,y)=\sum_{s}\frac{\lambda_{s}}{2}[\overline{\psi}_{s}^{*}(x)-\braket{\overline{\psi}_{s}^{*}}][\overline{\psi}_{s}(y)-\braket{\overline{\psi}_{s}}].\label{eq:Q0spectralsum}
\end{equation}

A pair of equation (\ref{eq:P0psi}) and (\ref{eq:Q0phi}) form a
system for determining $\overline{\psi}_{s}$, $\overline{\varphi}_{s}$,
and $\lambda_{s}$. This can be conveniently done by extending linear
variables $x$, $y,\ldots$ into the complex plane and using the expressions
for the kernels of $\hat{P}_{0}$ and $\hat{Q}_{0}$ treated as operator
densities,
\begin{align}
P_{0}(x-y,x_{0}) & =\frac{1}{2\pi}\partial_{x}\frac{x-y}{(x-y)^{2}+x_{0}^{2}},\label{eq:P0(x-y)}\\
Q_{0}(x-y,x_{0}) & =-\frac{1}{4\pi}\ln\{4[(x-y)^{2}+x_{0}^{2}]\},\label{eq:Q0(x-y)}
\end{align}
where $x_{0}>0$ which is introduced to handle singularities for $x=y$,
assuming the limit $x_{0}\to0$ in the end. Following Ref.~\citep{PhysRevD.98.125008},
let us introduce the function
\begin{equation}
\omega(z)=\ln\frac{1+z}{1-z}\label{eq:omega(z)}
\end{equation}
which is uniquely defined on the complex plane of $z$ with the cut
$-1\leq z\leq1$. Its values on the upper side of the cut $\omega_{+}(z)$
and on the lower side $\omega_{-}(z)$ are related as
\begin{equation}
\omega_{+}(z)-\omega_{-}(z)=-2\pi i,\label{eq:omega jump}
\end{equation}
and $\omega_{+}(z)$ on the upper side of the cut (when $z=x=\cos v$
with $0<v<\pi$) reads
\begin{equation}
\omega_{+}(x)\equiv\omega(x)=\ln\frac{1+x}{1-x}=-2\ln\tan\frac{v}{2}.\label{eq:omega(u)}
\end{equation}
The solution of the equations (\ref{eq:P0psi}) and (\ref{eq:Q0phi})
is based on the following integral (see Appendix~\ref{App:Integral})
\begin{widetext}

\begin{equation}
\lim_{x_{0}\to0}\fintop_{-1}^{+1}dyF(x-y,x_{0})f_{+}(y,s)=\pi i\coth(\pi s)\left[f_{+}(x,s)-\frac{1}{\cosh(\pi s)}\right]\label{eq:K integral}
\end{equation}
with 
\[
F(x-y,x_{0})=\frac{x-y}{(x-y)^{2}+x_{0}^{2}}
\]
and $f_{+}(x,s)$ being the value of the function $f(z,s)=\exp[-is\omega(z)]$
on the upper side of the cut, $f_{+}(x,s)=\exp[-is\omega_{+}(x)]$.

Then for $P_{0}(x-y,x_{0})=-(1/2\pi)F(x-y,x_{0})$, see Eq.~(\ref{eq:P0(x-y)}),
in the limit $x_{0}\to0$, one finds
\begin{equation}
\left(\hat{P}_{0}f_{+}\right)(x,s)=\frac{1}{2\pi}\partial_{x}\left\{ \pi i\coth(\pi s)\left[f_{+}(x,s)-\frac{1}{\cosh(\pi s)}\right]\right\} =\frac{1}{2}\coth(\pi s)i\partial_{x}f_{+}(x,s)\label{eq:P0f+}
\end{equation}
and
\begin{equation}
\left(\hat{Q}_{0}i\partial_{x}f_{+}\right)(x,s)=\frac{1}{2}\coth(\pi s)\left[f_{+}(x,s)-\frac{1}{\cosh(\pi s)}\right],\label{eq:Q0df+}
\end{equation}
\end{widetext}, in performing the integration by part, we have neglected
the boundary contributions at $x=\pm1$ because they rapidly oscillated
with $s$ and, therefore, do not contribute to the final integrals
over $s$ \citep{PhysRevD.98.125008}. Note also that the function
$f_{+}(x,s)$, being rewritten in the angular coordinate $v$ ($x=\cos v$),
takes the form $f_{+}(v,s)=\exp[2is\ln\tan(v/2)]$ and satisfies
\[
\intop_{0}^{\pi}dv\left[f_{+}(v,s)-\frac{1}{\cosh(\pi s)}\right]=0.
\]
As a result, Eqs.~(\ref{eq:P0f+}) and (\ref{eq:Q0df+}) suggest
the following solution of equations (\ref{eq:P0psi}) and (\ref{eq:Q0phi}):
\begin{equation}
\overline{\psi}_{s}(x)=\frac{1}{\sqrt{|s|}}f_{+}(x,s)=\frac{1}{\sqrt{|s|}}\exp[-is\omega(x)],\label{eq:psis(x)}
\end{equation}
\begin{align}
\overline{\varphi}_{s}(x) & =i\frac{\mathrm{sign}(s)}{\sqrt{|s|}}\partial_{x}f_{+}(x,s)=i\mathrm{sign}(s)\partial_{x}\overline{\psi}_{s}(x)\label{eq:phis(x)}\\
 & =\frac{2\sqrt{|s|}}{1-x^{2}}\exp[-is\omega(x)],\nonumber 
\end{align}
where $\mathrm{sign}(s)$ is the sign function, and
\begin{equation}
\lambda_{s}=\coth(\pi|s|)\label{eq:lambdas}
\end{equation}
resulting in 
\begin{equation}
\epsilon_{s}=2\pi|s|.\label{eq:epsilons}
\end{equation}
The normalization coefficient $(|s|)^{-1/2}$ in (\ref{eq:psis(x)})
is chosen according to (\ref{eq:complete set conditions}) with $\delta_{s,s'}$
and $\sum_{s}$ being replaced with $2\pi\delta(s-s')$ and $\int_{-\infty}^{\infty}ds/2\pi$,
respectively {[}integration over $x$ in (\ref{eq:complete set conditions})
is from $-1$ to $1${]}.

In the angular variable $v$, the functions $\overline{\psi}_{s}(v)$
and $\overline{\varphi}_{s}(v)$ are
\begin{align}
\overline{\psi}_{s}(v) & =\frac{1}{\sqrt{|s|}}\exp(2is\ln\tan\frac{v}{2}),\label{eq:psis(v)}\\
\overline{\varphi}_{s}(v) & =-i\mathrm{sign}(s)\partial_{v}\overline{\psi}_{s}(v)\label{eq:phis(v)}\\
 & =2\sqrt{|s|}\frac{1}{\sin v}\exp(2is\ln\tan\frac{v}{2}).\nonumber 
\end{align}
Note that $\overline{\psi}_{s}(x=\cos v)=\overline{\psi}_{s}(v)$
but $\overline{\varphi}_{s}(x=\cos v)=\overline{\varphi}_{s}(v)/\sin v$,
and the functions (\ref{eq:psis(v)}) and (\ref{eq:phis(v)}) satisfy
the following completeness and biorthogonality conditions
\begin{equation}
\intop_{-\infty}^{\infty}\frac{ds}{2\pi}\overline{\psi}_{s}(v)\overline{\varphi}_{s}^{*}(w)=\delta(v-w),\label{eq:completeness angular}
\end{equation}
\begin{equation}
\intop_{0}^{\pi}dv\overline{\psi}_{s}^{*}(v)\overline{\varphi}_{s'}(v)=2\pi\delta(s-s')\label{eq:orthogonality angular}
\end{equation}
{[}In performing the integral over the angular variable, one can use
the fact that $d\ln\tan(v/2)=\sin^{-1}vdv$.{]} The functions $\psi_{s}(v)$
{[}or $\psi_{s}(x)${]} and $\varphi_{s}(v)$ {[}or $\varphi_{s}(x)${]}
can be obtained from $\overline{\psi}_{s}$ and $\overline{\varphi}_{s}$
by rescaling: 
\begin{align*}
\psi_{s}(v) & =\sqrt{\frac{2}{\lambda_{s}}}\overline{\psi}_{s}=\sqrt{\frac{2}{|s|\lambda_{s}}}\exp(2is\ln\tan\frac{v}{2}),\\
\varphi_{s}(v) & =\sqrt{\frac{\lambda_{s}}{2}}\overline{\varphi}_{s}=-i\mathrm{sign}(s)\frac{\lambda_{s}}{2}\partial_{v}\psi_{s}\\
 & =\sqrt{2|s|\lambda_{s}}(\sin v)^{-1}\exp(2is\ln\tan\frac{v}{2}).
\end{align*}
It follows that 
\[
\psi_{s=0}(v)=\sqrt{2\pi},
\]
\[
\varphi_{s=0}(v)=\sqrt{2/\pi}(\sin v)^{-1},
\]
and 
\[
\int_{0}^{\pi}dv\varphi_{s}(v)=\sqrt{2\pi}\delta(s),
\]
 which justifies the condition (\ref{eq:W20rhs}).

As a check, let us calculate the kernels of $\hat{P}_{0}$, $\hat{P}_{0}^{-1}$,
$\hat{Q}_{0}^{-1}$ and $\hat{Q}_{0}$ using their spectral decomposition.
For $\hat{P}_{0}$ in angular variables one obtains\begin{widetext}
\begin{align}
P_{0}(v_{1},v_{2}) & =\intop_{-\infty}^{\infty}\frac{ds}{2\pi}\varphi_{s}^{*}(v_{1})\varphi_{s}(v_{2})=\intop_{-\infty}^{\infty}\frac{ds}{2\pi}\frac{1}{2}\coth(\pi|s|)\overline{\varphi}_{s}^{*}(v_{1})\overline{\varphi}_{s}(v_{2})\nonumber \\
 & =\frac{1}{\sin v_{2}\sin v_{1}}\intop_{-\infty}^{\infty}\frac{ds}{2\pi}2s\coth(\pi s)\exp[is\Omega(v_{1},v_{2})],\label{eq:P0 angular}
\end{align}
\end{widetext}where
\begin{equation}
\Omega(v_{1},v_{2})=2\left(\ln\tan\frac{v_{2}}{2}-\ln\tan\frac{v_{1}}{2}\right).\label{eq:Omega}
\end{equation}
The integral over $s$ in (\ref{eq:P0 angular}) can be performed
by closing contour of integration at infinity in the complex upper
half plane for $\Omega>0$ or in the lower one for $\Omega<0$, and
then using the residue theorem with the result {[}see Eq.~(\ref{eq:Ks0final}){]}
\[
P_{0}(v_{1},v_{2})=-\frac{1}{2\pi}\frac{\sin v_{1}\sin v_{2}}{(\cos v_{1}-\cos v_{2})^{2}}.
\]
In a similar way,
\begin{align*}
(P_{0}^{-1})(v_{1},v_{2}) & =\intop_{-\infty}^{\infty}\frac{ds}{2\pi}\psi_{s}^{*}(v_{1})\psi_{s}(v_{2})\\
 & =\intop_{-\infty}^{\infty}\frac{ds}{2\pi}\frac{2\tanh(\pi s)}{s}\exp[is\Omega(v_{1},v_{2})]\\
 & =\frac{1}{\pi}\ln\frac{1-\cos(v_{1}-v_{2})}{1-\cos(v_{1}+v_{2})},
\end{align*}
and
\begin{align*}
(Q_{0}^{-1})(v_{1},v_{2}) & =\intop_{-\infty}^{\infty}\frac{ds}{2\pi}[2\tanh(\pi|s|)]^{2}\varphi_{s}^{*}(v_{1})\varphi_{s}(v_{2})\\
 & =\partial_{v_{1}}\partial_{v_{2}}\intop_{-\infty}^{\infty}\frac{ds}{2\pi}\frac{2\tanh(\pi s)}{s}\exp[is\Omega(v_{1},v_{2})]\\
 & =\frac{1}{\pi}\partial_{v_{1}}\partial_{v_{2}}\ln\frac{1-\cos(v_{1}-v_{2})}{1-\cos(v_{1}+v_{2})},
\end{align*}
see (\ref{eq:Kc0final}). 

Following Eq.~(\ref{eq:Q0spectralsum}), the expression for the kernel
$Q_{0}(v_{2},v_{1})$ is
\begin{equation}
Q_{0}(v_{1},v_{2})=\intop_{-\infty}^{\infty}\frac{ds}{2\pi}\frac{\coth(\pi|s|)}{2}[\overline{\psi}_{s}^{*}(v_{1})-\braket{\overline{\psi}_{s}}][\overline{\psi}_{s}(v_{2})-\braket{\overline{\psi}_{s}}],\label{eq:Q0spectral}
\end{equation}
where
\begin{equation}
\braket{\overline{\psi}_{s}}=\frac{1}{\pi}\intop_{0}^{\pi}du\overline{\psi}_{s}(u)=\frac{1}{\sqrt{|s|}}\frac{1}{\cosh(\pi s)}.\label{eq:psi bar averaged}
\end{equation}
The result of the integration (see Appendix \ref{App:Calculation of Q0}
for details) reads\begin{widetext}
\begin{align*}
Q_{0}(v_{1},v_{2}) & =\intop_{-\infty}^{\infty}\frac{ds}{2\pi}\frac{\coth(\pi s)}{2s}\left[\exp(-2is\ln\tan\frac{v_{1}}{2})-\frac{1}{\cosh(\pi s)}\right]\left[\exp(2is\ln\tan\frac{v_{2}}{2})-\frac{1}{\cosh(\pi s)}\right],\\
 & =-\frac{1}{4\pi}\ln[4(\cos v_{1}-\cos v_{2})^{2}]
\end{align*}
\end{widetext}which coincides with Eq.~(\ref{eq:Kc0-1}). Note that
$Q_{0}(v_{1},v_{2})$ satisfies the condition $\left(\hat{Q}_{0}\hat{Q}_{0}^{-1}\right)(v_{1},v_{2})=\left(\hat{Q}_{0}^{-1}\hat{Q}_{0}\right)(v_{1},v_{2})=\delta(v_{1}-v_{2})-1/\pi$. 

Finally, let us calculate the kernels in the entanglement Hamiltonian
following Eqs.~(\ref{eq:HEpi general}) and (\ref{eq:HEphi general})
which still hold despite of the modified spectral decomposition (\ref{eq:Q0spectral})
of $\hat{Q}_{0}$ because it does not enter the expression for the
reduced density matrix. The results are
\begin{align*}
2\pi H_{E\pi}(u,v) & =\intop_{-\infty}^{\infty}\frac{ds}{2\pi}\pi\exp[2is(\ln\tan\frac{u}{2}-\ln\tan\frac{v}{2})]\\
 & =\frac{\pi}{2}\sin(u)\delta(u-v)
\end{align*}
or 
\[
2\pi H_{E\pi}(x,y)=\frac{\pi}{2}(1-x^{2})\delta(x-y),
\]
and
\[
2\pi H_{E\phi}(u,v)=\frac{\pi}{2}\partial_{u}\partial_{v}[\sin(u)\delta(u-v)]
\]
or
\[
2\pi H_{E\phi}(u,v)=\frac{\pi}{2}\partial_{x}\partial_{y}[(1-x^{2})\delta(x-y)].
\]
 We thus obtain the well-known result
\begin{align*}
\hat{H}_{E} & =\frac{1}{4}\intop_{0}^{\pi}du\sin(u)[\hat{\pi}_{u}^{2}+(\partial_{u}\hat{\phi}_{u})^{2}]\\
 & =\frac{1}{4}\intop_{-1}^{1}dx(1-x^{2})[\hat{\pi}_{x}^{2}+(\partial_{x}\hat{\phi}_{x})^{2}].
\end{align*}

\section{Small $M$ limit .\label{sec:limit Mto0.}}

The general result (\ref{eq:RDM general}) can conveniently be used
for the analysis of the small mass limit, $M\to0$. For this purpose,
one can use the known results for the Mathieu and associated Mathieu
functions which enter Eqs.~(\ref{eq:Kc}) and (\ref{eq:Ks}) for
the kernels in (\ref{eq:RDM general}) (see some details in Appendix~\ref{App:Mathieu functions}).
One finds that the leading correction to the $M=0$ result comes from
the kernel $K_{c}$ and has the form

\[
\delta K_{c}(u,v)=-\frac{1}{2\pi L}\sim\frac{1}{\ln(1/M)}
\]
and is coordinate independent. Here $L=\ln(\sqrt{q}/2)+\gamma=\ln(M/4)+\gamma$
with $\gamma$ being the Euler constant. The subleading corrections
start with terms $\sim M^{2}\ln(1/M)$ and will not be considered. 

In the leading order, therefore, the kernels $K_{c}$and $K_{s}$
in Eq.~(\ref{eq:RDM general}) are
\begin{align*}
K_{c}(u,v) & \approx K_{c}^{(0)}(u-v)+\delta K_{c}(u,v)=K_{c}^{(0)}(u-v)+\frac{1}{\pi\Lambda},\\
K_{s}(u,v) & \approx K_{s}^{(0)}(u,v),
\end{align*}
where $K_{c}^{(0)}(u-v)$ and $K_{s}^{(0)}(u,v)$ are kernels for
$M=0$ and $\Lambda=-2L=\ln(4/q)-2\gamma$. Within this approximation,
equations for the functions $\psi_{s}(x)$ and $\varphi_{s}(x)$ read
\begin{align}
\{\hat{Q}_{0}^{-1}+[\frac{1}{\pi\Lambda}]-\hat{P}_{0}(\frac{2}{\lambda_{s}})^{2}\}\psi_{s} & =0,\label{eq:equation for psi}\\
\varphi_{s} & =\hat{P}_{0}\psi_{s},\label{eq:phi via psi}
\end{align}
where $\hat{Q}_{0}^{-1}$ and $\hat{P}_{0}$ correspond to $M=0$,
and $[1/\pi\Lambda]$ denotes an operator with the coordinate independent
kernel, $[1/\pi\Lambda](x,y)=(\pi\Lambda)^{-1}$, such that 
\[
([1/\pi\Lambda]\psi_{s})(u)=\frac{1}{\pi\Lambda}\int_{0}^{\pi}dv\psi_{s}(v)
\]
is $u$-independent. One can now use the small parameter $1/(\pi\Lambda)\ll1$
to treat Eq.~(\ref{eq:equation for psi}) perturbatively and find
the eigenvalues $\lambda_{s}$ and the eigenfunctions $\psi_{s}$
{[}and then $\varphi_{s}$ from Eq.~(\ref{eq:phi via psi}){]} in
terms of $M=0$ eigenvalue $\lambda_{s}^{(0)}$ and eigenfunctions
$\psi_{s}^{(0)}$ and $\varphi_{s}^{(0)}$. The corresponding calculations
are presented in Appendix~\ref{App:Perturbative calculations}, and
here we give an alternative derivation based on the spectral decomposition
of the kernels
\begin{align}
Q^{-1}(u,v) & =\sum_{s}\left[\frac{2}{\lambda_{s}}\right]^{2}\varphi_{s}^{*}(u)\varphi_{s}(v),\label{eq:Spectraldecomp}\\
P(u,v) & =\sum_{s}\psi_{s}^{*}(u)\psi_{s}(v)\label{eq:SpectraldecompP}
\end{align}
and 
\begin{equation}
2\pi H_{E\pi}(u,v)=\frac{1}{2}\sum\epsilon_{s}\frac{\lambda_{s}}{2}\psi_{s}^{*}(u)\psi_{s}(v)=\frac{1}{2}(\hat{L}\hat{Q})(u,v),\label{eq:Hpi}
\end{equation}
\begin{equation}
2\pi H_{E\phi}(u,v)=\frac{1}{2}\sum\epsilon_{s}\frac{2}{\lambda_{s}}\varphi_{s}^{*}(u)\varphi_{s}(v)=\frac{1}{2}(\hat{P}\hat{L})(u,v),\label{eq:Hphi}
\end{equation}
see Eq.~(\ref{eq:EntHamc}). Here
\begin{equation}
\hat{L}=2\hat{g}\ln\frac{1+\hat{g}}{1-\hat{g}}=\ln\left[\frac{1+\hat{g}}{1-\hat{g}}\right]2\hat{g}\label{eq:L through g}
\end{equation}
with $\hat{g}^{2}=(\hat{Q}\hat{P})^{-1}/4=\hat{P}{}^{-1}\hat{Q}{}^{-1}/4$,
see, for example, \citep{Peschel_2009,IngoPeschel_2003} for different
but equivalent expressions. The equations (\ref{eq:Hpi}) and (\ref{eq:Hphi})
follow from the result for the kernel of the operator $\hat{L}$, 

\begin{align*}
L(u,v) & =\sum_{s}\frac{2}{\lambda_{s}}\ln\frac{\lambda_{s}+1}{\lambda_{s}-1}\psi_{s}^{*}(u)\varphi_{s}(v)\\
 & =\sum_{s}\epsilon_{s}\frac{2}{\lambda_{s}}\psi_{s}^{*}(u)\varphi_{s}(v),
\end{align*}
which one obtains after substituting the expression $g(u,v)=\sum_{s}\lambda_{s}^{-1}\psi_{s}^{*}(u)\varphi_{s}(v)$
for the kernel of $\hat{g}$ in Eq.~(\ref{eq:L through g}). 

To proceed, let us represent $\hat{L}$ in the following form (the
eigenvalues $\lambda_{s}^{-2}$ of $\hat{g}^{2}$ satisfy the condition
$0\leq\lambda_{s}^{-1}<1$)\begin{widetext}
\begin{equation}
\hat{L}=4\hat{g}^{2}\intop_{1}^{\infty}\frac{dx}{x^{2}-\hat{g}^{2}}=4\hat{g}^{2}\intop_{1}^{\infty}dx\intop_{0}^{\infty}d\tau\exp[-\tau(x^{2}-\hat{g}^{2})]=4\hat{g}^{2}\intop_{0}^{\infty}d\tau f(\tau)\exp[\tau\hat{g}^{2}]\label{eq:IntegralL}
\end{equation}
\end{widetext}with 
\[
f(\tau)=\intop_{1}^{\infty}dx\exp[-\tau x^{2}]=\frac{1}{2}\sqrt{\frac{\pi}{\tau}}\mathrm{erfc}(\sqrt{\tau}),
\]
where $\mathrm{erfc}(z)$ is the complementary error function. As
it follows from the definition, the function $f(\tau)$ has the following
property
\begin{equation}
\intop_{0}^{\infty}d\tau f(\tau)\exp[\tau/\lambda_{s}^{2}]=\frac{\lambda_{s}}{2}\ln\frac{\lambda_{s}+1}{\lambda_{s}-1}=\epsilon_{s}\frac{\lambda_{s}}{2}\label{eq:fproperty}
\end{equation}
for $1<\lambda_{s}<\infty$.

The representation (\ref{eq:IntegralL}) allows to find the change
of $\hat{L}$ when $\hat{g}^{2}\to\hat{g}^{2}+\hat{\delta}$. To the
leading order in $\hat{\delta}$, one has\begin{widetext}
\begin{equation}
\delta\hat{L}=4\hat{\delta}\intop_{0}^{\infty}d\tau f(\tau)\exp[\tau\hat{g}^{2}]+4\hat{g}^{2}\intop_{0}^{\infty}d\tau f(\tau)\intop_{0}^{\tau}d\tau'\exp[(\tau-\tau')\hat{g}^{2}]\hat{\delta}\exp[\tau'\hat{g}^{2}]\label{eq:deltaL1}
\end{equation}
or, alternatively,
\begin{equation}
\delta\hat{L}=\intop_{0}^{\infty}d\tau f(\tau)\exp[\tau\hat{g}^{2}]4\hat{\delta}+\intop_{0}^{\infty}d\tau f(\tau)\intop_{0}^{\tau}d\tau'\exp[(\tau-\tau')\hat{g}^{2}]\hat{\delta}\exp[\tau'\hat{g}^{2}]4\hat{g}^{2}.\label{eq:deltaL2}
\end{equation}
The kernel of $\delta\hat{L}$ can be found by using the spectral
decomposition of $\hat{g}^{2}$, biorthogonality of the functions
$\psi_{s}$, $\varphi_{s}$, and Eq.~(\ref{eq:fproperty}) with the
result
\[
\delta L(u,v)=\sum_{s_{1},s_{2}}\frac{\epsilon_{s_{1}}(2/\lambda_{s_{1}})-\epsilon_{s_{2}}(2/\lambda_{s_{2}})}{(2/\lambda_{s_{1}})^{2}-(2/\lambda_{s_{2}})^{2}}\left(\varphi_{s_{1}}4\hat{\delta}\psi_{s_{2}}^{*}\right)\psi_{s_{1}}^{*}(x)\varphi_{s_{2}}(y),
\]
\end{widetext}where
\[
\left(\varphi_{s_{1}}4\hat{\delta}\psi_{s_{2}}^{*}\right)=4\iintop_{0}^{\pi}dudv\varphi_{s_{1}}(u)\delta(u,v)\psi_{s_{2}}^{*}(v).
\]

We can now apply the above expression to our case with $4\hat{\delta}=\hat{P}_{0}{}^{-1}\delta\hat{Q}{}^{-1}$
to obtain
\[
\left(\varphi_{s_{1}}4\hat{\delta}\psi_{s_{2}}^{*}\right)=\frac{1}{\pi\Lambda}\Psi_{s_{1}}\Psi_{s_{2}}^{*},
\]
where we define
\begin{equation}
\Psi_{s}=\int_{0}^{\pi}du\psi_{s}(u)=\sqrt{\frac{2}{|s|\lambda_{s}}}\frac{\pi}{\cosh(\pi s)}\label{eq:Psis}
\end{equation}
and omit referring of all the eigenfunctions and eigenvalues to the
unperturbed system. For $\delta\hat{L}(x,y)$ we then get\begin{widetext}
\[
\delta L(u,v)=\frac{1}{\pi\Lambda}\sum_{s_{1},s_{2}}\frac{\epsilon_{s_{1}}(2/\lambda_{s_{1}})-\epsilon_{s_{2}}(2/\lambda_{s_{2}})}{(2/\lambda_{s_{1}})^{2}-(2/\lambda_{s_{2}})^{2}}\Psi_{s_{1}}\Psi_{s_{2}}^{*}\psi_{s_{1}}^{*}(u)\varphi_{s_{2}}(v),
\]
where the term with $s_{1}=s_{2}$ should be understood as derivative
$\partial[\epsilon_{s}(2/\lambda_{s})]/\partial[(2/\lambda_{s})^{2}]$.
As a result, the correction to the field part of the entanglement
Hamiltonian, Eq.~(\ref{eq:Hphi}), is 
\begin{equation}
2\pi\delta H_{E\phi}(u,v)=\frac{1}{2}(\hat{P}_{0}\delta\hat{L})(u,v)=\frac{1}{2\pi\Lambda}\sum_{s_{1},s_{2}}\frac{\epsilon_{s_{1}}(2/\lambda_{s_{1}})-\epsilon_{s_{2}}(2/\lambda_{s_{2}})}{(2/\lambda_{s_{1}})^{2}-(2/\lambda_{s_{2}})^{2}}\Psi_{s_{1}}\Psi_{s_{2}}^{*}\varphi_{s_{1}}^{*}(u)\varphi_{s_{2}}(v).\label{eq:deltaHphi}
\end{equation}
In a similar way, for the correction to the field part of the entanglement
Hamiltonian, Eq.~(\ref{eq:Hpi}) we obtain {[}here the second form
of $\hat{L}$ in (\ref{eq:L through g}) is more convenient{]}
\begin{align}
2\pi\delta H_{E\pi}(u,v) & =\frac{1}{2}\delta(\hat{L}\hat{Q})(u,v)=\frac{1}{2}\left[\intop_{0}^{\infty}d\tau f(\tau)\intop_{0}^{\tau}d\tau'\exp[(\tau-\tau')\hat{g}^{2}]\hat{\delta}\exp[\tau'\hat{g}^{2}]\hat{P_{0}}^{-1}\right](u,v)\label{eq:deltaHpi}\\
 & =\frac{1}{2\pi\Lambda}\sum_{s_{1},s_{2}}\frac{\epsilon_{s_{1}}(\lambda_{s_{1}}/2)-\epsilon_{s_{2}}(\lambda_{s_{2}}/2)}{(2/\lambda_{s_{1}})^{2}-(2/\lambda_{s_{2}})^{2}}\Psi_{s_{1}}\Psi_{s_{2}}^{*}\psi_{s_{1}}^{*}(u)\psi_{s_{2}}(v),\nonumber 
\end{align}
where, as before, the term with $s_{1}=s_{2}$ is understood as derivative
$\partial[\epsilon_{s}(\lambda_{s}/2)]/\partial[(2/\lambda_{s})^{2}]$.
Note that, despite of the fact that $\varphi_{s}(u)\sim\partial_{u}\psi_{s}(u)$,
as it follows from (\ref{eq:deltaHphi}) and (\ref{eq:deltaHpi}),
$\delta H_{E\phi}(u,v)\neq\partial_{u}\partial_{v}\delta H_{E\pi}(u,v)$.

Using the above results for $\delta H_{E\phi}(u,v)$ and $\delta H_{E\pi}(u,v)$
together with Eqs.~(\ref{eq:lambdas})-(\ref{eq:phis(v)}) and (\ref{eq:Psis})
one obtains
\begin{align}
2\pi\delta H_{E\pi}(u,v) & =\frac{1}{2\pi\Lambda}\iintop_{-\infty}^{\infty}\frac{ds_{1}ds_{2}}{(2\pi)^{2}}\frac{4}{\lambda_{s_{1}}\lambda_{s_{2}}}\frac{1}{|s_{1}||s_{2}|}\frac{\pi|s_{1}|\lambda_{s_{1}}-\pi|s_{2}|\lambda_{s_{2}}}{(2/\lambda_{s_{1}})^{2}-(2/\lambda_{s_{2}})^{2}}\frac{\pi^{2}}{\cosh(\pi s_{1})\cosh(\pi s_{2})}\label{eq:deltaHpi integral}\\
 & \times\exp[-2is_{1}\ln\tan\frac{u}{2}+2is_{2}\ln\tan\frac{v}{2}]\nonumber 
\end{align}
and
\begin{align}
2\pi\delta H_{E\phi}(u,v) & =\frac{1}{2\pi\Lambda}\partial_{u}\partial_{v}\iintop_{-\infty}^{\infty}\frac{ds_{1}ds_{2}}{(2\pi)^{2}}\frac{1}{s_{1}s_{2}}\frac{4\pi|s_{1}|/\lambda_{s_{1}}-4\pi|s_{2}|/\lambda_{s_{2}}}{(2/\lambda_{s_{1}})^{2}-(2/\lambda_{s_{2}})^{2}}\frac{\pi^{2}}{\cosh(\pi s_{1})\cosh(\pi s_{2})}\label{eq:deltaHphi integral}\\
 & \times\exp[-2is_{1}\ln\tan\frac{u}{2}+2is_{s}\ln\tan\frac{v}{2}]\nonumber \\
 & =\frac{1}{2\pi\Lambda}\frac{4}{\sin(u)\sin(v)}\iintop_{-\infty}^{\infty}\frac{ds_{1}ds_{2}}{(2\pi)^{2}}\frac{4\pi|s_{1}|/\lambda_{s_{1}}-4\pi|s_{2}|/\lambda_{s_{2}}}{(2/\lambda_{s_{1}})^{2}-(2/\lambda_{s_{2}})^{2}}\frac{\pi^{2}}{\cosh(\pi s_{1})\cosh(\pi s_{2})}\nonumber \\
 & \times\exp[-2is_{1}\ln\tan\frac{u}{2}+2is_{2}\ln\tan\frac{v}{2}].\nonumber 
\end{align}

We start calculations with the integral (\ref{eq:deltaHpi integral})
which can be rewritten as
\[
2\pi\delta H_{E\pi}(u,v)=\frac{\pi^{2}}{4\Lambda}\iintop_{-\infty}^{\infty}\frac{ds_{1}ds_{2}}{(2\pi)^{2}}\frac{1}{s_{1}s_{2}}\left\{ \frac{s_{1}-s_{2}}{\sinh[\pi(s_{1}-s_{2})]}-\frac{s_{1}+s_{2}}{\sinh[\pi(s_{1}+s_{2})]}\right\} \exp[-2is_{1}\ln\tan\frac{u}{2}+2is_{2}\ln\tan\frac{v}{2}].
\]
After introducing $s_{\pm}=s_{1}\pm s_{2}$, one obtains
\[
2\pi\delta H_{E\pi}(u,v)=\frac{\pi^{2}}{2\Lambda}\iintop_{-\infty}^{\infty}\frac{ds_{+}ds_{-}}{(2\pi)^{2}}\frac{1}{s_{+}^{2}-s_{-}^{2}}\left\{ \frac{s_{-}}{\sinh(\pi s_{-})}-\frac{s_{+}}{\sinh(\pi s_{+})}\right\} \exp[is_{+}\Omega_{-}(u,v)+is_{+}\Omega_{-}(u,v)],
\]
where $\Omega_{\pm}(u,v)=\left|\ln\tan\frac{u}{2}\pm\ln\tan\frac{v}{2}\right|$
and we change the integration variables $s_{\pm}\to-s_{\pm}$. (That
the modulus in the definition of $\Omega_{\pm}(u,v)$ is because negative
values of $\ln\tan\frac{u}{2}\pm\ln\tan\frac{v}{2}$ can be compensated
by the substitution $s_{\pm}\to-s_{\pm}$.) We first perform the integration
over $s_{+}$ by closing the integration contour in the upper half-plane
and using the residue theorem with the result (note that the integrand
is regular for $s_{+}=\pm s_{-}$ and only the poles at $s_{+}=in$
with positive integer $n$ contribute)
\[
\intop_{-\infty}^{\infty}\frac{ds_{+}}{2\pi}\frac{1}{s_{+}^{2}-s_{-}^{2}}\left\{ \frac{s_{-}}{\sinh(\pi s_{-})}-\frac{s_{+}}{\sinh(\pi s_{+})}\right\} \exp[is_{+}\Omega_{-}(u,v)]=-\frac{1}{\pi}\sum_{n=1}^{\infty}(-1)^{n}\frac{n}{n^{2}+s_{-}^{2}}\exp[-n\Omega_{-}(u,v)].
\]
The integration over $s_{-}$ then gives
\begin{align*}
2\pi\delta H_{E\pi}(u,v) & =\frac{\pi}{4\Lambda}\sum_{n=1}^{\infty}(-1)^{n+1}\exp\{-n[\Omega_{-}(u,v)+\Omega_{+}(u,v)]\}\\
 & =\frac{\pi}{4\Lambda}\frac{1}{e^{[\Omega_{+}(u,v)+\Omega_{-}(u,v)]}+1}=\frac{\pi}{8\Lambda}[1-\max(|\cos u|,|\cos v|)],
\end{align*}
such that the correction to the momentum part of the entanglement
Hamiltonian is
\[
\delta\hat{H}_{E\pi}=\frac{1}{16\Lambda}\iintop_{0}^{\pi}dudv[1-\max(|\cos u|,|\cos v|)]\hat{\pi}_{u}\hat{\pi}_{v}=\frac{1}{16\Lambda}\iintop_{-1}^{1}dxdy[1-\max(|x|,|y|)]\hat{\pi}_{x}\hat{\pi}_{y}.
\]

The calculation of the $2\pi\delta H_{E\phi}(u,v)$ is similar: After
changing to variables $s_{\pm}$, one has
\[
2\pi\delta H_{E\phi}(u,v)=\frac{\pi^{2}}{2\Lambda}\frac{1}{\sin(u)\sin(v)}\iintop_{-\infty}^{\infty}\frac{ds_{+}ds_{-}}{(2\pi)^{2}}\left\{ \frac{s_{-}}{\sinh(\pi s_{-})}+\frac{s_{+}}{\sinh(\pi s_{+})}\right\} \exp[-is_{+}\Omega_{-}(u,v)-is_{-}\Omega_{+}(u,v)],
\]
and, using the integral
\[
\intop_{-\infty}^{\infty}\frac{ds}{2\pi}\frac{s}{\sinh(\pi s)}\exp(2isL)=\frac{1}{4\pi}\frac{1}{\cosh^{2}L},
\]
 the result is
\begin{align*}
2\pi\delta H_{E\phi}(u,v) & =\frac{\pi}{8\Lambda}\frac{\sin u}{\sin v}\left[\delta(\ln\tan\frac{u}{2}+\ln\tan\frac{v}{2})+\delta(\ln\tan\frac{u}{2}-\ln\tan\frac{v}{2})\right]\\
 & =\frac{\pi}{8\Lambda}\sin u\left[\delta(u+v-\pi)+\delta(u-v)\right].
\end{align*}
The correction to the field part of the entanglement Hamiltonian thus
reads
\[
\delta\hat{H}_{E\phi}=\frac{1}{16\Lambda}\intop_{0}^{\pi}du\sin(u)[\hat{\phi}_{u}^{2}+\hat{\phi}_{u}\hat{\phi}_{\pi-u}]=\frac{1}{16\Lambda}\intop_{-1}^{1}dx[\hat{\phi}_{x}^{2}+\hat{\phi}_{x}\hat{\phi}_{-x}]=\frac{1}{32\Lambda}\intop_{-1}^{1}dx(\hat{\phi}_{x}+\hat{\phi}_{-x})^{2}.
\]
It is worth to mention that the above expression can be rewritten
as
\begin{align*}
\delta\hat{H}_{E\phi} & =\frac{1}{16\Lambda}\left\{ \iintop_{-1}^{1}dxdy\{\partial_{x}\partial_{y}[1-\max(|x|,|y|)]\}\hat{\phi}_{x}\hat{\phi}_{y}+2\intop_{-1}^{1}dx\hat{\phi}_{x}\hat{\phi}_{-x}\right\} \\
 & =\frac{1}{16\Lambda}\left\{ \iintop_{-1}^{1}dxdy[1-\max(|x|,|y|)]\partial_{x}\hat{\phi}_{x}\partial_{y}\hat{\phi}_{y}+2\intop_{-1}^{1}dx\hat{\phi}_{x}\hat{\phi}_{-x}\right\} ,
\end{align*}
where the boundary contribution in the last line was neglected.

Finally, the leading order correction to the entanglement Hamiltonian
are
\begin{align*}
\delta\hat{H}_{E} & =\frac{1}{16\Lambda}\left\{ \iintop_{-1}^{1}dxdy[1-\max(|x|,|y|)]\hat{\pi}_{x}\hat{\pi}_{y}+\intop_{-1}^{1}dx[\hat{\phi}_{x}^{2}+\hat{\phi}_{x}\hat{\phi}_{-x}]\right\} \\
 & =\frac{1}{16\Lambda}\left\{ \iintop_{-1}^{1}dxdy[1-\max(|x|,|y|)](\hat{\pi}_{x}\hat{\pi}_{y}+\partial_{x}\hat{\phi}_{x}\partial_{y}\hat{\phi}_{y})+2\intop_{-1}^{1}dx\hat{\phi}_{x}\hat{\phi}_{-x}\right\} .
\end{align*}
\end{widetext}This expression contains couplings across the entire
interval, as well as ``antidiagonal'' couplings between points which
are symmetric with respect to the interval's center. Such types of
couplings were also found earlier for a free fermionic field \citep{PhysRevD.95.065005}
and for a chain of harmonic oscillators \citep{PhysRevD.95.065005,Eisler_2020}.
The coupling across the interval is zero at the ends of the interval
and has a form of a rectangular pyramid with the apex at $x=y=0$.
An unusual feature of the antidiagonal coupling is that it does not
vanish at the ends of the interval, which is in contrast to the fermionic
case in Ref.~\citep{PhysRevD.95.065005}. This difference can be
attributed to a more singular behavior of the eigenfunctions $\overline{\varphi}_{s}(x)$
close to the interval ends,$\overline{\varphi}_{s}(x)\sim(1-x^{2})^{-1}$,
see Eq.~\ref{eq:phis(x)}, as compared to the fermionic case where
the analogous eigenfunctions $\psi_{s}(x)$ behave as $\psi_{s}(x)\sim(1-x^{2})^{-1/2}$,
see Eq.~(A6) in Ref.~\citep{PhysRevD.95.065005}. As a result, the
factor $1-x^{2}$ which is present in the correction term for fermions,
disappears for bosons.

\section{Conclusion and discussion\label{sec:Discussion-and-Conclusions}}

Based on the exact solution for the reduced density matrix on an interval
for a $1+1$ dimensional scalar field in the ground state, the entanglement
Hamiltonian is discussed in the cases of a small and a large mass
of the field. In the small-mass case, the found corrections to the
zero-mass result are inversely proportional to the logarithm of the
mass (multiplied with the length of the interval). The corrections
contain two different type of couplings: The coupling between pairs
of points across the interval with a position-dependent strength which
is zero at the end points of the interval, and the coupling between
two points which are symmetric relative to the center of the interval.
An unusual feature of the latter antidiagonal coupling is the its
strength remains constant across the entire interval including the
end points.

The analysis of the large-mass case supports the picture of two weakly
coupled edges in the sense that the behavior of the local terms in
the entanglement Hamiltonian close to each interval end is similar
to that in a half-infinite interval, namely, the entanglement Hamiltonian
density, as a function of the distance from the corresponding boundary,
is proportional to the Hamiltonian density and grows linearly. More
precisely, a local operator with the density which is the Hamiltonian
density multiplied by a triangular function of position on the interval
(with zero at the end points) commutes with the reduced density matrix
up to order $M^{-1/2}$. This result implies that the structure of
the entanglement eigenfunctions $\psi_{s}$ and $\varphi_{s}$ is
strongly affected by the mass even for $s\to\infty$, and the analytic
structure of the local temperature is more complicated than that in
the massless case where it has two-pole structure \citep{PhysRevD.96.105019}
resulting in its parabolic profile. The importance of the mass for
all values of $s$can be argued as follows: When the mass term is
ignored, the equation for the function $S$ which determines the eigenfunctions
$\psi_{s}$ and $\varphi_{s}$, see Ref.~\citep{PhysRevD.96.105019},
acquires conformal properties of the massless case, which lead to
the massless solutions after using the conformal mapping of a finite
cut on the half-infinite one. However, under such mapping, the mass
term becomes too singular to be treated perturbatively, even for large
$s$, which contradicts the initial assumption of the mass-irrelevance
in the large $s$ limit. This also indicates that the parameter $s$
which enters the boundary condition for the field $S$, cannot be
associated with a kind of ``energy'' in the field equation where
the high-energy solutions are similar in the massive and massless
cases. A complete understanding of the effect of the mass on the structure
of the entanglement eigenfunctions for all values of $s$ requires
additional studies.
\begin{acknowledgments}
The authors acknowledge helpful discussions with M. Dalmonte and T.
Zache. This work is supported by the European Union’s Horizon Europe
research and innovation program under Grant Agreement No. 101113690
(PASQuanS2.1).
\end{acknowledgments}

\appendix

\section{Mathieu functions\label{App:Mathieu functions}}

This Appendix contains a summary of the properties of Mathieu and
associated Mathieu functions, which are used in the main text. Here
we mainly follow Refs.~\citep{mclachlan1947theory} and \citep{meixnerandSchaefke1954},
and our notations are from \citep{mclachlan1947theory}. We start
with the trigonometric series for the periodic Mathieu functions $\mathrm{ce}_{m}(z,-q)$
and $\mathrm{se}_{m}(z,-q)$ which satisfy Eq.~(\ref{eq:gstandard})
with the substitution $q\to-q=-(M/2)^{2}$ and have eigenvalues $a_{m}$
and $b_{m}$ indicated after each series:\begin{widetext}
\begin{equation}
\mathrm{ce}_{2n}(z,-q)=(-1)^{n}\mathrm{ce}_{2n}(\frac{\pi}{2}-z,q)=(-1)^{n}\sum_{r=0}^{\infty}(-1)^{r}A_{2r}^{(2n)}\cos(2rz),\:a_{2n},\label{eq:ce2n}
\end{equation}
\begin{equation}
\mathrm{ce}_{2n+1}(z,-q)=(-1)^{n}\mathrm{se}_{2n+1}(\frac{\pi}{2}-z,q)=(-1)^{n}\sum_{r=0}^{\infty}(-1)^{r}B_{2r+1}^{(2n+1)}\cos[(2r+1)z],\:b_{2n+1},\label{eq:ce2n+1}
\end{equation}
\begin{equation}
\mathrm{se}_{2n+1}(z,-q)=(-1)^{n}\mathrm{ce}_{2n+1}(\frac{\pi}{2}-z,q)=(-1)^{n}\sum_{r=0}^{\infty}(-1)^{r}A_{2r+1}^{(2n+1)}\sin[(2r+1)z],\:a_{2n+1},\label{eq:se2n+1}
\end{equation}
\begin{equation}
\mathrm{se}_{2n+2}(z,-q)=(-1)^{n}\mathrm{se}_{2n+2}(\frac{\pi}{2}-z,q)=(-1)^{n}\sum_{r=0}^{\infty}(-1)^{r}A_{2r+2}^{(2n+2)}\sin[(2r+2)z],\:b_{2n+2},\label{eq:se2n+2}
\end{equation}
\end{widetext}where $n=0,1,\ldots$ and the coefficients $A_{s}^{(m)}$,
$B_{s}^{(m)}$ satisfy three-point recursive relations, see~\citep{mclachlan1947theory},
which determine their, as well as of the eigenvalues $a_{m}$and $b_{m}$,
dependence on $q$. These functions are normalized to $\pi$,
\[
\intop_{0}^{2\pi}[\mathrm{ce/se}_{m}(z,-q)]^{2}dz=\pi,
\]
and orthogonal to each other on the interval $[0,2\pi]$. Their periodicity
and transformation properties under $z\to-z$ and $z\to\pi-z$ can
be seen from the above expressions. We also mention that the set of
functions \{$\mathrm{ce}_{m}(z,-q),\mathrm{se}_{m}(z,-q)$\} is complete
on the interval $0<v<2\pi$, while each of the set\{$\mathrm{ce}_{m}(z,-q)$\}
and \{$\mathrm{se}_{m}(z,-q)$\} is complete on the interval $0<v<\pi$.

For $q=0$, the only nonzero coefficients are $A_{0}^{(0)}=1/\sqrt{2}$,
$A_{m}^{(m)}=B_{m}^{(m)}=1$ for $m\geq1,$and the eigenvalues $a_{m}^{2}=b_{m}^{2}=m^{2}$.
For small $q$, the leading behavior of $A_{s}^{(m)}$, $B_{s}^{(m)}$
is ($r\geq0$, $m>0$)
\[
A_{m+2r}^{(m)},\:B_{m+2r}^{(m)}\approx(-1)^{r}\frac{m!}{r!(m+r)!}\left[\frac{q}{4}\right]^{r}
\]
and
\[
A_{m-2r}^{(m)},\:B_{m-2r}^{(m)}\approx(-1)^{r}\frac{(m-r-1)!}{r!(m-1)!}\left[\frac{q}{4}\right]^{r}.
\]
This gives the small $q$ behavior of the Mathieu functions used in
Section \ref{sec:limit Mto0.}:
\[
\mathrm{ce}_{0}(z,-q)\approx\frac{1}{\sqrt{2}}\{1+\frac{1}{2}q\cos2z+\ldots\},
\]
\[
\mathrm{ce}_{1}(z,-q)\approx\cos z+\frac{1}{8}q\cos3z+\ldots,
\]
\[
\mathrm{se}_{1}(z,-q)\approx\sin z+\frac{1}{8}q\sin3z+\ldots,
\]
and for $m\geq2$
\begin{align*}
\mathrm{ce}_{m}(z,-q) & \approx\cos mz\\
 & +\frac{1}{4}q\left[\frac{\cos(m+2)z}{m+1}-\frac{\cos(m-2)z}{m-1}\right]+\ldots,
\end{align*}
\begin{align*}
\mathrm{se}_{m}(z,-q) & \approx\sin mz\\
 & +\frac{1}{4}q\left[\frac{\sin(m+2)z}{m+1}-\frac{\sin(m-2)z}{m-1}\right]+\ldots,
\end{align*}
where the omitted terms contain higher powers of $q=(M/2)^{2}.$ 

The corresponding modified Mathieu functions $\mathrm{Fek}_{m}(u,-q)$
and $\mathrm{Gek}_{m}(u,-q)$ are given by the following absolutely
and monotonously converging (including point $z=0$) series\begin{widetext}

\begin{equation}
\mathrm{Fek}_{2n}(z,-q)=\frac{p'_{2n}}{\pi A_{0}^{(2n)}}\sum_{r=0}^{\infty}A_{2r}^{(2n)}\mathrm{I}_{r}(v_{1})\mathrm{K}_{r}(v_{2}),\label{eq:Fek2n}
\end{equation}
\begin{equation}
\mathrm{Fek}_{2n+1}(z,-q)=\frac{s'_{2n+1}}{\pi B_{1}^{(2n+1)}}\sum_{r=0}^{\infty}B_{2r+1}^{(2n+1)}[\mathrm{I}_{r}(v_{1})\mathrm{K}_{r+1}(v_{2})-\mathrm{I}_{r+1}(v_{1})\mathrm{K}_{r}(v_{2})],\label{eq:Fek2n+1}
\end{equation}
\begin{equation}
\mathrm{Gek}_{2n+1}(z,-q)=\frac{p'_{2n+1}}{\pi A_{1}^{(2n+1)}}\sum_{r=0}^{\infty}A_{2r+1}^{(2n+1)}[\mathrm{I}_{r}(v_{1})\mathrm{K}_{r+1}(v_{2})+\mathrm{I}_{r+1}(v_{1})\mathrm{K}_{r}(v_{2})],\label{eq:Gek2n+1}
\end{equation}
\begin{equation}
\mathrm{Gek}_{2n+2}(z,-q)=\frac{s'_{2n+2}}{\pi B_{2}^{(2n+2)}}\sum_{r=0}^{\infty}B_{2r+2}^{(2n+2)}[\mathrm{I}_{r}(v_{1})\mathrm{K}_{r+2}(v_{2})-\mathrm{I}_{r+2}(v_{1})\mathrm{K}_{r}(v_{2})],\label{eq:Gen2n+2}
\end{equation}
\end{widetext}where $v_{1}=(M/2)\exp(-z)$, $v_{2}=(M/2)\exp(z)$,
$\mathrm{I}_{r}(v)$ and $\mathrm{K}_{r}(v)$ are modified Bessel
functions, and 
\[
p'_{2n}=\frac{(-1)^{n}}{A_{0}^{(2n)}}\mathrm{ce}_{2n}(0,q)\mathrm{ce}_{2n}(\frac{\pi}{2},q),
\]
\[
p'_{2n+1}=\frac{2(-1)^{n+1}}{MA_{1}^{(2n+1)}}\mathrm{ce}_{2n+1}(0,q)\mathrm{ce}'_{2n+1}(\frac{\pi}{2},q),
\]
\[
s'_{2n+1}=\frac{2(-1)^{n}}{MB_{1}^{(2n+1)}}\mathrm{se}'_{2n+1}(0,q)\mathrm{se}{}_{2n+1}(\frac{\pi}{2},q),
\]
\[
s'_{2n+2}=\frac{4(-1)^{n+1}}{M^{2}B_{2}^{(2n+2)}}\mathrm{se}'_{2n+2}(0,q)\mathrm{se}'_{2n+2}(\frac{\pi}{2},q),
\]
with $\mathrm{ce}'_{2n+1}(z,q)=d\mathrm{ce}{}_{2n+1}(z,q)/dz,$etc.,
being the derivatives with respect to $z$.

The asymptotics of the functions (\ref{eq:Fek2n})-(\ref{eq:Gen2n+2})
for large positive $z$ has the form
\[
\mathrm{Fek/Gek}_{m}(z,-q)\sim\frac{C}{\sqrt{2\pi v_{2}}}\exp(-v_{2}),
\]
where the constant $C$ coincides with $c'_{m}$ or $p'_{m}$ which
appears in the definition of the functions.

For small $q$ and not very large $z$ such that $\sqrt{q}\exp(z)\ll1$,
one can use properties of the modified Bessel functions to obtain
the leading behavior of the modified Mathieu functions (\ref{eq:Fek2n})-(\ref{eq:Gen2n+2}):
\begin{align*}
\mathrm{Fek}_{0}(z,-q) & \to-\frac{1}{\pi\sqrt{2}}[z+\ln\frac{\sqrt{q}}{2}+\gamma][1+q\cosh2z],\\
\mathrm{Fek/Gek}_{m}(z,-q) & \to2^{2m-2}(m-1)!m!\frac{1}{\pi q^{m}}\exp(-mz)
\end{align*}
for $m\geq1$, and for their log-derivatives in the kernels $K_{c}$
and $K_{s}$
\begin{align*}
\frac{\mathrm{Fek}'_{0}(0,-q)}{\mathrm{Fek}_{0}(0,-q)} & \approx\frac{1}{\ln\frac{\sqrt{q}}{2}+\gamma}(1+q+\ldots),\\
\frac{\mathrm{Fek}'_{1}(0,-q)}{\mathrm{Fek}_{1}(0,-q)} & \approx-[1-q(\ln\frac{\sqrt{q}}{2}+\gamma+1/4)+\ldots],\\
\frac{\mathrm{Gek}'_{1}(0,-q)}{\mathrm{Gek}_{1}(0,-q)} & \approx-[1-q(\ln\frac{\sqrt{q}}{2}+\gamma-3/4)+\ldots],
\end{align*}
and
\[
\frac{\mathrm{Fek'}_{m}(0,-q)}{\mathrm{Fek}_{m}(0,-q)}\approx\frac{\mathrm{Gek'}_{m}(0,-q)}{\mathrm{Gek}_{m}(0,-q)}\approx-m(1+\frac{q}{m^{2}-1}+\ldots)
\]
when $m\geq2$ and the omitted terms are of the order of $q^{2}\ln q$.

The results for small $q$ mentioned in Section \ref{sec:limit Mto0.}
directly follow from the above expressions.

For $M\to\infty$ ($q\to\infty$), the asymptotic behavior of the
eigenvalues and Mathieu functions can be found in Refs.~\citep{BG}
and \citep{meixnerandSchaefke1954}. Namely, one has 

\begin{equation}
b_{m+1}-a_{m}=\frac{1}{m!}2^{4m+5}q^{(m+3/2)/2}e^{-4q^{1/2}}[1+O(q^{-1/2})]\label{eq:am bm+1}
\end{equation}
and 
\begin{equation}
\frac{a_{m}+2q}{4q^{1/2}}=m+\frac{1}{2}+\frac{\gamma_{(1)m}}{4q^{1/2}}+O(q^{-1})\label{eq:am large q}
\end{equation}
with 
\begin{equation}
\gamma_{(1)m}=-\frac{1}{2}\left[(m+\frac{1}{2})^{2}+\frac{1}{4}\right].\label{eq:gamma1}
\end{equation}

For the corresponding Mathieu functions {[}up to terms $O(q^{-1/2})${]}
one has the following relations
\begin{align}
ce_{m+1}(z,-q) & =ce_{m}(z,-q)+O(q^{-1}),\label{eq:cem cem+1 large q around 0}\\
se_{m+1}(z,-q) & =se_{m}(z,-q)+O(q^{-1})\label{eq:sem sem+1 large q around 0}
\end{align}
for $z\ll1$ and 
\begin{align}
ce_{m+1}(z,-q) & =-ce_{m}(z,-q)+O(q^{-1}),\label{eq:cem cem+1 large q around pi}\\
se_{m+1}(z,-q) & =-se_{m}(z,-q)+O(q^{-1})\label{eq:sem sem+1 large q around pi}
\end{align}
for $\pi-z\ll1$. 

The asymptotic of the Mathieu functions with $m=2n$ (when $a=a_{2n}\approx b_{2n+1}$)
reads\begin{widetext}
\begin{equation}
ce_{2n}(z,-q)=(-1)^{n}C_{2n}\left[\mathrm{D}_{2n}(\alpha)+\frac{1}{4q^{1/2}}y_{(1)n}(\alpha)+O(q^{-1})\right],\label{eq:ce2n large q}
\end{equation}
where

\begin{equation}
y_{(1)n}(\alpha)=-\frac{1}{16}\mathrm{D}_{2n+4}(\alpha)-\frac{1}{4}\mathrm{D}_{2n+2}(\alpha)-\frac{n(2n-1)}{2}\mathrm{D}_{2n-2}(\alpha)+\frac{n(n-1)(2n-1)(2n-3)}{4}\mathrm{D}_{2n-4}(\alpha)\label{eq:y1n}
\end{equation}
and
\begin{equation}
C_{2n}=\left[\frac{\pi q^{1/2}}{2}\right]^{1/4}\frac{1}{\sqrt{(2n)!}}\left[1-\frac{4n+1}{16q^{1/2}}+O(q^{-1})\right].\label{eq:C2n large q}
\end{equation}
\end{widetext}The functions $ce_{2n+1}(z,-q)$ are then given by
(\ref{eq:cem cem+1 large q around 0}) and (\ref{eq:cem cem+1 large q around pi}).
In the above formulae, $\alpha=2q^{1/4}\sin z$ and $\mathrm{D}_{m}(\alpha)$
is the parabolic cylinder functions $\mathrm{D}_{p}(\alpha)$ with
non-negative integer order

\begin{align*}
\mathrm{D}_{m}(\alpha) & =(-1)^{m}\exp(\alpha^{2}/4)\frac{d^{m}}{d\alpha^{m}}\exp(-\alpha^{2}/2)\\
 & =2^{-m/2}\exp(-\alpha^{2}/4)\mathrm{H}_{m}(\alpha/\sqrt{2}),
\end{align*}
where $\mathrm{H}_{m}(x)$ is the Hermite polynomial, which are normalized
as
\begin{equation}
\intop_{0}^{\infty}d\alpha\mathrm{D}_{m}^{2}=\sqrt{\frac{\pi}{2}}(m)!.\label{eq:normDp}
\end{equation}

The results for $m=2n+1$ (when $a=a_{2n+1}\approx b_{2n+2}$) are\begin{widetext}
\begin{equation}
se_{2n+1}(z,-q)=(-1)^{n}C_{2n+1}\left[\mathrm{D}_{2n+1}(\alpha)+\frac{1}{4q^{1/2}}\overline{y}_{(1)n}(\alpha)+O(q^{-1})\right]\label{eq:se2n+1 large q}
\end{equation}
with 
\begin{equation}
\overline{y}_{(1)n}(\alpha)=-\frac{1}{16}\mathrm{D}_{2n+5}(\alpha)-\frac{1}{4}\mathrm{D}_{2n+3}(\alpha)-\frac{n(2n+1)}{2}\mathrm{D}_{2n-1}(\alpha)+\frac{n(n-1)(2n+1)(2n-1)}{4}\mathrm{D}_{2n-3}(\alpha)\label{eq:ybar1n}
\end{equation}
and
\begin{equation}
C_{2n+1}=\left[\frac{\pi q^{1/2}}{2}\right]^{1/4}\frac{1}{\sqrt{(2n+1)!}}\left[1-\frac{4n+3}{16q^{1/2}}+O(q^{-1})\right].\label{eq:C2n+1 large q}
\end{equation}
\end{widetext}The functions $se_{2n+2}(z,-q)$ are obtained from
(\ref{eq:sem sem+1 large q around 0}) and (\ref{eq:sem sem+1 large q around pi}).

To obtain the asymptotic form of the corresponding modified Mathieu
function $\mathrm{Fek/Gek}_{m}(u,-q)$ for $q\to\infty$, one can
introduce the new variable $\zeta=2q^{1/4}\sinh u$ and rewrite Eq.~(\ref{eq:f})
in the form
\[
\left[\frac{d^{2}}{d\zeta^{2}}-\frac{a+2q}{4q^{1/2}}-\frac{\zeta^{2}}{4}\right]f=-\frac{1}{4q^{1/2}}\left[\zeta^{2}\frac{d^{2}}{d\zeta^{2}}+\zeta\frac{d}{d\zeta}\right]f
\]
for $f=f(\zeta)$, which, together with the results for the eigenvalues,
Eqs.~(\ref{eq:am bm+1})-(\ref{eq:gamma1}), allows an expansion
of the solutions in powers of $q^{-1/2}$. In the leading order, the
right-hand-side of this equation can be neglected, and one finds for
$a=a_{m}$
\[
f_{(0)m}(\zeta)=\mathrm{D}_{-(m+1)}(\zeta),
\]
where the parabolic cylinder functions $\mathrm{D}_{-\nu}(z)$ for
$\nu>0$ has the following integral representation
\begin{equation}
\mathrm{D}_{-\nu}(z)=\frac{1}{\Gamma(\nu)}\exp(-\frac{z^{2}}{4})\intop_{0}^{\infty}t^{\nu-1}\exp(-zt-\frac{t^{2}}{2})dt\label{eq:D-nu}
\end{equation}
with $\Gamma(\nu)$ being the Euler gamma function.

The next orders can be found using the well-known functional relations
for the parabolic cylinder function 
\begin{equation}
\frac{d}{dz}\mathrm{D}_{p}(z)=\frac{1}{2}[p\mathrm{D}_{p-1}(z)-\mathrm{D}_{p+1}(z)]\label{eq:derivative Dp}
\end{equation}
and
\begin{equation}
z\mathrm{D}_{p}(z)=p\mathrm{D}_{p-1}(z)+\mathrm{D}_{p+1}(z),\label{eq:zDp}
\end{equation}
together with the equation for $\mathrm{D}_{-p}(z)$,
\[
\left[\frac{d^{2}}{dz^{2}}+\frac{1}{2}-\frac{z^{2}}{4}\right]\mathrm{D}_{-p}(z)=p\mathrm{D}_{-p}(z).
\]
Omitting unimportant for our purpose {[}see Eqs.~(\ref{eq:Kc}) and
(\ref{eq:Ks}){]} prefactors, one then obtains
\begin{align}
\mathrm{Fek}_{2n}(u,-q) & =\mathrm{D}_{-2n-1}(\zeta)+\frac{1}{4q^{1/2}}f_{(1)n}(\zeta)\label{eq:Fek2n large q}\\
 & +O(q^{-1}),\nonumber \\
\mathrm{Fek}_{2n+1}(u,-q) & =\mathrm{Fek}_{2n}(u,-q)+O(q^{-1}),\label{eq:Fek2n+1 large q}
\end{align}
 for $a=a_{2n}\approx b_{2n+1}$ with\begin{widetext}
\begin{equation}
f_{(1)n}(\zeta)=\frac{1}{16}\mathrm{D}_{-2n+3}(\alpha)+\frac{1}{4}\mathrm{D}_{-2n+1}(\alpha)+\frac{(n+1)(2n+1)}{2}\mathrm{D}_{-2n-3}(\alpha)-\frac{(n+1)(n+2)(2n+1)(2n+3)}{4}\mathrm{D}_{-2n-5}(\alpha).\label{eq:f1n}
\end{equation}
The results for $a=a_{2n+1}\approx b_{2n+2}$ are
\begin{align}
\mathrm{Gek}_{2n+1}(u,-q) & =\mathrm{D}_{-2n-2}(\zeta)+\frac{1}{4q^{1/2}}\overline{f}_{(1)n}(\zeta)+O(q^{-1}),\label{eq:Gek2n+1 large q}\\
\mathrm{Gek}_{2n+2}(u,-q) & =\mathrm{Gek}_{2n+1}(u,-q)+O(q^{-1}),\label{eq:Gek2n+2 large q}
\end{align}
with
\begin{equation}
\overline{f}_{(1)n}(\zeta)=\frac{1}{16}\mathrm{D}_{-2n+2}(\alpha)+\frac{1}{4}\mathrm{D}_{-2n}(\alpha)+\frac{(n+1)(2n+3)}{2}\mathrm{D}_{-2n-4}(\alpha)-\frac{(n+1)(n+2)(2n+3)(2n+5)}{4}\mathrm{D}_{-2n-6}(\alpha).\label{eq:fbar1n}
\end{equation}
\end{widetext}In the above formulae, unimportant for our purpose
{[}see Eqs.~(\ref{eq:Kc}) and (\ref{eq:Ks}){]} ``normalization''
constants are omitted.

With the above asymptotic expansions, one can calculate the log-derivatives
of the modified Mathieu functions, which appear in kernels $K_{c}$
and $K_{s}$, Eqs.~(\ref{eq:Kc}) and (\ref{eq:Ks}), as well as
their products in Eqs.~(\ref{eq:commutator matrix element}) and
(\ref{eq:commutator matrix element 1}). After using the expressions
\[
\mathrm{D}_{-\nu}(0)=\frac{\sqrt{\pi}}{2^{\nu/2}\Gamma(\frac{\nu+1}{2})},\,\mathrm{D'}_{-\nu}(0)=-\frac{\sqrt{\pi}}{2^{(\nu-1)/2}\Gamma(\nu/2)}
\]
which follow from (\ref{eq:D-nu}), one obtains (note that $d/du|_{u=0}=2q^{1/4}d/d\zeta|_{\zeta=0}$)\begin{widetext}
\begin{equation}
\frac{\mathrm{Fek'}_{2n}(0,-q)}{\mathrm{Fek}_{2n}(0,-q)}=\frac{\mathrm{Fek'}_{2n+1}(0,-q)}{\mathrm{Fek}_{2n+1}(0,-q)}=-2\sqrt{2}q^{1/4}\frac{\Gamma(n+1)}{\Gamma(n+1/2)}\left[1-\frac{n+1/4}{8q^{1/2}}+O(q^{-1})\right],\label{eq:logderFek}
\end{equation}
\begin{equation}
\frac{\mathrm{Gek'}_{2n+1}(0,-q)}{\mathrm{Gek}_{2n+1}(0,-q)}=\frac{\mathrm{Gek'}_{2n+2}(0,-q)}{\mathrm{Gek}_{2n+2}(0,-q)}=-2\sqrt{2}q^{1/4}\frac{\Gamma(n+3/2)}{\Gamma(n+1)}\left[1-\frac{n+3/4}{8q^{1/2}}+O(q^{-1})\right],\label{eq:logderGek}
\end{equation}
and
\[
\frac{\mathrm{Fek'}_{2n_{1}}(0,-q)}{\mathrm{Fek}_{2n_{1}}(0,-q)}\frac{\mathrm{Gek'}_{2n_{2}+1}(0,-q)}{\mathrm{Gek}_{2n_{2}+1}(0,-q)}=\frac{\mathrm{Fek'}_{2n_{1}+1}(0,-q)}{\mathrm{Fek}_{2n_{1}+1}(0,-q)}\frac{\mathrm{Gek'}_{2n_{2}+2}(0,-q)}{\mathrm{Gek}_{2n_{2}+2}(0,-q)}
\]
\begin{equation}
=8q^{1/2}\frac{\Gamma(n_{1}+1)}{\Gamma(n_{1}+1/2)}\frac{\Gamma(n_{2}+3/2)}{\Gamma(n_{2}+1)}\left[1-\frac{1}{8q^{1/2}}(n_{1}+n_{2}+1)+O(q^{-1})\right].\label{eq:logderFekGek}
\end{equation}
\end{widetext}

\section{Reduced density matrix and correlation functions\label{App:Reduced-density-matrix}.}

In this Appendix we present a direct derivation of the reduced density
matrix for an interval starting from the ground state wave function
of the entire system. This derivation shows, in particular, the connection
between the reduced density matrix of the subsystem and the correlation
functions in the entire system. To simplify the consideration and
expressions, let us consider field $\hat{\phi}_{x}$ and momentum
$\hat{\pi}_{x}$ operators being defined on a circle with a circumference
$L\to\infty$ (an interval $x\in[0,L]$ with periodic boundary conditions).
This geometry preserves the translational symmetry which results in
a zero mode in the massless case. In the field representation, for
a state of the field $\ket{\phi(x)}$ one has
\begin{equation}
\hat{\phi}_{x}\ket{\phi(x)}=\phi(x)\ket{\phi(x)},\:\hat{\pi}_{x}\ket{\phi(x)}=-i\frac{\delta}{\delta\phi(x)}\ket{\phi(x)}.\label{eq:pi and phi}
\end{equation}
In this representation, the ground state $\ket{G}$ of the Hamiltonian
(\ref{eq:H}) has the following wave function
\begin{equation}
\left\langle \phi(x)|G\right\rangle \sim\exp\{-\frac{1}{4}\int dxdy\phi(x)K(x-y)\phi(y)\},\label{eq:GS wave function}
\end{equation}
where (and in expressions below) the limits of integration are from
$0$ to $L$. As it can be easily checked, the kernel $K(x-y)$ is
related to the momentum-momentum correlation function
\begin{equation}
\bra{G}\hat{\pi}_{x}\hat{\pi}_{y}\ket{G}\equiv P(x-y)=\frac{1}{4}K(x-y)\label{eq:momentu correlations}
\end{equation}
and its inverse to the field-field correlation function 
\begin{equation}
\bra{G}\hat{\phi}_{x}\hat{\phi}_{y}\ket{G}\equiv Q(x-y)=(K)^{-1}(x-y),\label{eq:field correlations}
\end{equation}
where $Q(x-y)$ satisfies
\[
\int dzQ(x-z)K(z-y)=\delta(x-y).
\]

Let us now consider an interval $\mathrm{I}_{l}$ of the length $2l$
inside $L$ such that $L=\bar{L}+2l$. The fields defined in the interval
will be denoted as $\phi_{l}$ and those in $\bar{L}$ as $\bar{\phi}$.
With this notation, the ground state wave function reads\begin{widetext}
\begin{align*}
\left\langle \phi(x)|G\right\rangle  & \sim\exp\{-\frac{1}{4}[\int dxdy\bar{\phi}(x)K_{\bar{L}\bar{L}}(x-y)\bar{\phi}(y)+\int dxdy\phi_{l}(x)K_{ll}(x-y)\phi_{l}(y)\\
 & +\int dxdy\bar{\phi}(x)K_{\bar{L}l}(x-y)\phi_{l}(y)+\int dxdy\phi_{l}(x)K_{l\bar{L}}(x-y)\bar{\phi}(y)]\},
\end{align*}
where the regions of integration follow from the indices of the kernels
and fields.

To find the reduced density matrix for the interval, let us consider
two field configuration: $\phi_{1}(x)=\{\bar{\phi}(x),\phi_{l1}(x)\}$
and $\phi_{1}(x)=\{\bar{\phi}(x),\phi_{l2}(x)\}$, in which the (same)
first components are for $x\in\bar{L}$ and the (different) second
ones for $x\in\mathrm{I}_{l}$. The reduced density matrix $\bra{\phi_{l1}(x)}\hat{\rho}\ket{\phi_{l2}(y)}$
is then obtained from the density matrix for the entire system {[}we
use the fact that $K_{\bar{L}l}(x-y)=K_{l\bar{L}}(x-y)${]}
\begin{align*}
\left\langle \phi_{1}(x)|G\right\rangle \left\langle G|\phi_{2}(y)\right\rangle  & \sim\exp\{-\frac{1}{2}\int dxdy\bar{\phi}(x)K_{\bar{L}l}(x-y)\bar{\phi}(y)-\frac{1}{2}\int dxdy\bar{\phi}(x)K_{\bar{L}l}(x-y)[\phi_{l1}(y)+\phi_{l2}(y)]\\
 & -\frac{1}{4}\int dxdy\phi_{l1}(x)K_{ll}(x-y)\phi_{l1}(y)-\frac{1}{4}\int dxdy\phi_{l2}(x)K_{ll}(x-y)\phi_{l2}(y)]\}
\end{align*}
by integrating out the field $\bar{\phi}(x)$ in $\bar{L}$. This
results in
\begin{align*}
\bra{\phi_{l1}(x)}\hat{\rho}\ket{\phi_{l2}(y)} & \sim\exp\{-\frac{1}{8}\int dxdy\phi_{-}(x)K_{ll}(x-y)\phi_{-}(y)\\
 & -\frac{1}{8}\int dxdy\phi_{+}(x)[K_{ll}(x-y)-K_{l\bar{L}}(K_{\bar{L}\bar{L}})^{-1}K_{\bar{L}l}(x,y)]\phi_{+}(y)\},
\end{align*}
\end{widetext}where $\phi_{\pm}(x)=\phi_{l1}(x)\pm\phi_{l2}(x)$
and we use symbolic notation for the product of the operators given
by the kernels $K_{l\bar{L}}(x-y)$, $(K_{\bar{L}\bar{L}})^{-1}(x,y)$,
and $K_{\bar{L}l}(x-y)$. This product can be expressed through the
blocks of the kernel $G(x-y)$, see Eq.~(\ref{eq:field correlations}),
as
\[
K_{l\bar{L}}(K_{\bar{L}\bar{L}})^{-1}K_{\bar{L}l}=K_{ll}-(Q_{ll})^{-1},
\]
where again symbolic form is used with $(G_{ll})^{-1}$denoting the
kernel $(G_{ll})^{-1}(x,y)$ which is the inverse of $G_{ll}(x-y)$
in the region $\mathrm{I}_{l}$ such that 
\[
\intop_{\mathrm{I}_{l}}dz(G_{ll})^{-1}(x,z)G_{ll}(z-y)=\delta(x-y)
\]
 for $x,y\in\mathrm{I}_{l}$. We then finally obtain\begin{widetext}
\begin{equation}
\bra{\phi_{l1}(x)}\hat{\rho}\ket{\phi_{l2}(y)}\sim\exp\{-\frac{1}{8}\int dxdy[\phi_{+}(x)(Q_{ll})^{-1}(x,y)\phi_{+}(y)+\phi_{-}(x)K_{ll}(x-y)\phi_{-}(y)]\}.\label{eq:RDM}
\end{equation}
\end{widetext} can easily check that the field and momentum correlation
functions for the region $\mathrm{I}_{l}$ calculated with the use
of Eq.~(\ref{eq:RDM}) coincide with those for the entire system,
Eqs.~(\ref{eq:momentu correlations}) and (\ref{eq:field correlations}),
and comparison with (\ref{eq:RDM general}) gives the relations between
kernels $(Q_{ll})^{-1}(x,y),K_{ll}(x-y)$ and $K_{c}(v_{1},v_{2}),K_{s}(v_{1},v_{2})$.
Namely, $K_{c}(v_{1},v_{2})=\sin\nu_{1}\sin\nu_{2}(Q_{ll})^{-1}(\cos\nu_{1},\cos\nu_{2})$
and $K_{s}(v_{1},v_{2})=\sin\nu_{1}\sin\nu_{2}K_{ll}(\cos\nu_{1}-\cos\nu_{2})/4$.

\section{Calculation of the integral (\ref{eq:K integral}) from Section~\ref{sec:Diagonalization-for-M=00003D0}\label{App:Integral}}

for the function $f(z)=\exp[-is\omega(z)]$ which is analytic in the
complex plane except the cut where it has a multiplicative boundary
condition $f_{+}(z,s)=\exp(-2\pi s)f_{-}(z,s)$ with real $z$ ($-1<z<1$)
and $f_{\pm}(z,s)=\exp[-is\omega_{\pm}(z)]$,

Here we present details of the calculations of the integral (\ref{eq:K integral})
in Section~\ref{sec:Diagonalization-for-M=00003D0} with the function

\[
F(x-y,x_{0})=\frac{x-y}{(x-y)^{2}+x_{0}^{2}}
\]
which we consider as a function of complex variables $x$ and $y$.
As a function of $y$, it has two poles at $y_{\pm}=x\pm ix_{0}$
with residues $-1/2$ and a pole at infinity with the residue $1$.
The function $f(y,s)=\exp[-is\omega(y)]=\exp\{-is\ln[(1+y)/(1-y)]\}$
is analytic in the complex plane $y$ except the cut on the real axis
$-1\leq y\leq1$ where it has a multiplicative boundary condition
$f_{+}(y,s)=\exp(-2\pi s)f_{-}(y,s)$, where $f_{\pm}(y,s)=\exp[-is\omega_{\pm}(y)]$,
see (\ref{eq:omega jump}). When $y\to\infty$, the function $f(y)$
tends to the constant $f(y\to\infty)=\exp(\pi s)$.

Let us consider the following contour integral in the complex plane
\[
I=\ointop_{C_{R}}dyF(x-y,x_{0})f(y,s),
\]
where the integration is performed over the circle $C_{R}$ of a large
radius $R$ in the clockwise direction. We first calculate the integral
$I$ in the limit $R\to\infty$ when it is given by the residue at
infinity,
\[
I=2\pi i\exp(\pi s).
\]
We then deform the contour $C_{R}$ such that the integral is reduced
to a sum of three contour integrals: $I_{\pm}$ around the poles $p_{\pm}$
and $I_{\mathrm{cut}}$ around the cut. The integrals around the poles
are simply
\[
I_{\pm}=\pi if(x\pm ix_{0},s).
\]
For real $x\in(-1,1)$ and $x_{0}\to0$$,$the integrals $I_{\pm}$are
\[
I_{+}=\pi if_{+}(x,s),
\]
\[
I_{-}=\pi if_{-}(x,s)=\pi if_{+}(x,s)\exp(2\pi s),
\]
and the integral around the cut is

\[
I_{\mathrm{cut}}=(1-e^{2\pi s})\fintop_{-1}^{1}dyF(x-y,x_{0}=0)f_{+}(y,s),
\]
where the boundary condition was used. Combining the above results
together, one finds\begin{widetext}
\[
2\pi ie^{\pi s}=(1-e^{2\pi s})\fintop_{-1}^{1}dyF(x-y,x_{0}=0)f_{+}(y,s)+\pi i(1+e^{2\pi s})f_{+}(x,s),
\]
such that
\[
\fintop_{-1}^{1}dyF(x-y,x_{0}=0)f_{+}(y,s)=-\pi i\frac{1+e^{2\pi s}}{1-e^{2\pi s}}\left[f_{+}(x,s)-\frac{2e^{\pi s}}{1+e^{2\pi s}}\right]=\pi i\coth(\pi s)\left[f_{+}(x,s)-\frac{1}{\cosh(\pi s)}\right].
\]
\end{widetext}

\section{Calculation of $Q_{0}(u,v)$.\label{App:Calculation of Q0}}

Here we present the details of the calculation of the integral
\begin{align*}
Q_{0}(u,v) & =\intop_{-\infty}^{\infty}\frac{ds}{2\pi}\frac{\coth(\pi s)}{2s}\left[e^{-2is\ln\tan\frac{u}{2}}-\frac{1}{\cosh(\pi s)}\right]\\
 & \times\left[e^{2is\ln\tan\frac{v}{2}}-\frac{1}{\cosh(\pi s)}\right]
\end{align*}
for the kernel $Q_{0}(u,v$) in Section~\ref{sec:Diagonalization-for-M=00003D0}.
Let us consider the above integral as a contour one in the complex
plane of $s$ over the real axis. After noting that the integrand
is analytic at $s=0$, we deform the contour of the integration from
the real axis into the contour $C$ which consists of two parts of
the real axis $(-\infty,-\epsilon]$ and $[\epsilon,\infty)$ connected
by a half circle of radius $\epsilon$ in the upper half plane (here
$0<\epsilon\ll1$). The integral can then be splitted into four integrals,
$Q_{0}=I_{1}+I_{2}+I_{3}+I_{4}$, with

\[
I_{1}=\intop_{C}\frac{ds}{2\pi}\frac{\coth(\pi s)}{2s}\exp[is\Omega(u,v)],
\]
where $\Omega(u,v)=2[\ln\tan(v/2)-\ln\tan(u/2)$, see Eq.~(\ref{eq:Omega}),

\begin{align*}
I_{2} & =-\intop_{C}\frac{ds}{2\pi}\frac{\coth(\pi s)}{2s\cosh(\pi s)}\exp[2is\ln\tan\frac{v}{2}]\\
 & =-\intop_{C}\frac{ds}{2\pi}\frac{1}{2s\sinh(\pi s)}\exp[2is\ln\tan\frac{v}{2}],
\end{align*}
\begin{align*}
I_{3} & =-\intop_{C}\frac{ds}{2\pi}\frac{\coth(\pi s)}{2s\cosh(\pi s)}\exp[-2is\ln\tan\frac{u}{2}]\\
 & =-\intop_{C}\frac{ds}{2\pi}\frac{1}{2s\sinh(\pi s)}\exp[-2is\ln\tan\frac{u}{2}],
\end{align*}
and
\[
I_{4}=\intop_{C}\frac{ds}{2\pi}\frac{\coth(\pi s)}{2s\cosh^{2}(\pi s)}=\intop_{C}\frac{ds}{2\pi}\frac{1}{s\sinh(2\pi s)}.
\]

In calculating $I_{1}$ one can use the residue theorem by closing
the contour $C$ in the upper half plane for $\Omega>0$ with the
contribution of the first order poles at $s=ik$ with positive integer
$k$, or in the lower half plane for $\Omega<0$ when the first order
poles at $s=-ik$ and a second order pole at $s=0$ contribute. This
gives
\begin{align*}
I_{1} & =\frac{1}{4\pi}\{\Omega-\ln[2(\cosh\Omega-1)]\}\\
 & =\frac{1}{4\pi}\bigl\{2(\ln\tan\frac{v}{2}-\ln\tan\frac{u}{2})-\ln[2\frac{(\cos u-\cos v)}{\sin(u)\sin(v)}]^{2}\bigr\}.
\end{align*}
In a similar way one obtains
\[
I_{2}=\frac{1}{4\pi}\biggl[-2\ln\tan\frac{v}{2}+\ln\frac{4}{\sin^{2}(v)}\biggr]
\]
and\,
\[
I_{3}=\frac{1}{4\pi}\biggl[2\ln\tan\frac{u}{2}+\ln\frac{4}{\sin^{2}(u)}\biggr]
\]
such that
\[
I_{2}+I_{3}=\frac{1}{4\pi}\biggl[-\Omega+\ln\frac{16}{\sin^{2}(v)\sin^{2}(u)}\biggr].
\]
For $I_{4}$, after closing the contour in the upper half plane, one
gets
\[
I_{4}=-\frac{1}{4\pi}\ln16,
\]
such that the final result for $Q_{0}(u,v)$ is 
\[
Q_{0}(u,v)=-\frac{1}{4\pi}\ln[4(\cos u-\cos v)^{2}].
\]

\section{Perturbative calculations.\label{App:Perturbative calculations}}

Here we present an alternative derivation of the results (\ref{eq:deltaHphi})
and (\ref{eq:deltaHpi}) for the corrections to the entanglement Hamiltonian
in the small mass case using perturbative approach to the system (\ref{eq:equation for psi})
and (\ref{eq:phi via psi}). It is convenient to write this system
using the notation $\hat{\epsilon}$ for the perturbation operator
with the kernel $\varepsilon=1/\pi\Lambda$, see main text,
\begin{align}
\{\hat{Q}_{0}^{-1}+\hat{\epsilon}-\hat{P}_{0}(\frac{2}{\lambda_{s}})^{2}\}\psi_{s} & =0,\label{eq:Pert1}\\
\varphi_{s} & =\hat{P}_{0}\psi_{s}.\label{eq:Pert2}
\end{align}
For $\hat{\epsilon}=0$, the unperturbed eigenfunctions $\psi_{s}^{(0)}$,
$\varphi_{s}^{(0)}$, and the eigenvalues $\lambda_{s}^{(0)}$ satisfy
\begin{align}
\{\hat{Q}_{0}^{-1}-\hat{P}_{0}(\frac{2}{\lambda_{s}^{(0)}})^{2}\}\psi_{s}^{(0)} & =0,\label{eq:Pert10}\\
\varphi_{s}^{(0)} & =\hat{P}_{0}\psi_{s}^{(0)}.\label{eq:Pert20}
\end{align}
As it was mentioned in the main text, there is a zero mode for $M=0$
case, which we assign to $s=0$, i.e. $1/\lambda_{s}^{(0)}=0$, and
the corresponding eigenfunctions satisfy
\[
\hat{Q}_{0}^{-1}\psi_{0}^{(0)}=0,\:\varphi_{0}^{(0)}=\hat{P}_{0}\psi_{0}^{(0)}.
\]
 The goal is to express the solutions for $\psi_{s}$, $\varphi_{s}$,
and $\lambda_{s}$ in terms of $\psi_{s}^{(0)}$, $\varphi_{s}^{(0)}$
and $\lambda_{s}^{(0)}$ in the lowest order in $\hat{\epsilon}$.
From the spectral decomposition of the kernels (omitting coordinate
dependence) $\hat{P}=\sum_{s}\varphi_{s}\varphi_{s}^{*}$ ,$\hat{P}_{0}=\sum_{s}\varphi_{s}^{(0)}\varphi_{s}^{*(0)}$,
and the fact that $\hat{P}=\hat{P}_{0}$ to the first order in $\epsilon$,
it follows that the eigenfunctions $\varphi_{s}$ are obtained from
$\varphi_{s}^{(0)}$ by a (local) unitary rotations
\begin{equation}
\varphi_{s}=\sum_{p}S_{sp}\varphi_{p}^{(0)},\label{eq:phiSphi0}
\end{equation}
 where $S^{\dagger}=S^{-1}$. The same is also true for the eigenfunctions
$\psi_{s}$ because $\psi_{s}$ and $\varphi_{s}$ form a biorthogonal
system, 
\begin{equation}
\psi_{s}=\sum_{p}S_{sp}\psi_{p}^{(0)}.\label{eq:psiSpsi0}
\end{equation}
After using the spectral decomposition of $\hat{Q}_{0}^{-1}=\sum_{s}(2/\lambda_{s}^{(0)})^{2}\varphi_{s}^{*(0)}\varphi_{s}^{(0)}$,
Eq.~(\ref{eq:Pert1}) can be rewritten in the angular variables as
\begin{equation}
\intop_{0}^{\pi}du\bigl\{\sum_{p}[E_{p}^{(0)}-E_{s}]\varphi_{p}^{(0)}(v)\varphi_{p}^{*(0)}(u)\bigr\}\psi_{s}(u)=-\frac{1}{\pi\Lambda}\Psi_{s},\label{eq:Pert3}
\end{equation}
where $\Psi_{s}$ is given by Eq.~(\ref{eq:Psis}) and we define
\[
E_{p}=(2/\lambda_{p})^{2},\:E_{p}^{(0)}=(2/\lambda_{p}^{(0)})^{2}
\]
 (note that $E_{0}^{(0)}=0$). After multiplying Eq.~(\ref{eq:Pert3})
with $\psi_{q}^{*(0)}(v)$ and integrating over $v$, one obtains
with the use of Eq.~(\ref{eq:psiSpsi0}) that
\[
\left[E_{s}-E_{q}^{(0)}\right]S_{sq}=\frac{1}{\pi\Lambda}\sum_{p}\Psi_{q}^{*(0)}\Psi_{p}^{(0)}S_{sp},
\]
where $\Psi_{s}^{(0)}=\int_{0}^{\pi}dv\psi_{s}^{(0)}(v)$. The above
equation is similar to the one from the standard quantum mechanical
perturbation theory, and one gets to leading order in $1/\pi\Lambda$
\begin{align}
E_{s} & \approx E_{s}^{(0)}+E_{s}^{(1)}=E_{s}^{(0)}+\frac{1}{\pi\Lambda}\left|\Psi_{s}^{(0)}\right|^{2},\label{eq:Es}\\
S_{ss} & \approx1,\nonumber \\
S_{s\neq q} & \approx\frac{1}{\pi\Lambda}\frac{1}{E_{s}^{(0)}-E_{q}^{(0)}}\Psi_{q}^{*(0)}\Psi_{s}^{(0)},\nonumber 
\end{align}
such that\begin{widetext}
\begin{equation}
\psi_{s}(u)\approx\psi_{s}^{(0)}(u)+\psi_{s}^{(1)}(u)=\psi_{s}^{(0)}(u)+\frac{1}{\pi\Lambda}\sum_{q\neq s}\frac{\Psi_{q}^{*(0)}\Psi_{s}^{(0)}}{E_{s}^{(0)}-E_{q}^{(0)}}\psi_{q}^{(0)}(u)\label{eq:psispert}
\end{equation}
and
\begin{equation}
\varphi_{s}(u)\approx\varphi_{s}^{(0)}(u)+\varphi_{s}^{(1)}(u)=\varphi_{s}^{(0)}(u)+\frac{1}{\pi\Lambda}\sum_{q\neq s}\frac{\Psi_{q}^{*(0)}\Psi_{s}^{(0)}}{E_{s}^{(0)}-E_{q}^{(0)}}\varphi_{q}^{(0)}(u).\label{eq:phispert}
\end{equation}
\end{widetext}{[}As usual, in this approximation, the biorthogonality
condition and Eqs.~(\ref{eq:Pert1}) and (\ref{eq:Pert2}) are only
valid up to the first order in $(\pi\Lambda)^{-1}$.{]} 

With the above results, the first order correction to the entanglement
Hamiltonian, Eq.~(\ref{eq:EntHamc}), reads\begin{widetext}
\begin{align}
2\pi\delta H_{E} & =\frac{1}{2}\sum_{s}\left\{ E_{s}^{(1)}\frac{\partial}{\partial E_{s}^{(0)}}\left[\epsilon_{s}^{(0)}\frac{\lambda_{s}^{(0)}}{2}\right](\psi_{s}^{*(0)}\hat{\pi})(\psi_{s}^{(0)}\hat{\pi})+\epsilon_{s}^{(0)}\frac{\lambda_{s}^{(0)}}{2}\left[(\psi_{s}^{*(1)}\hat{\pi})(\psi_{s}^{(0)}\hat{\pi})+(\psi_{s}^{*(0)}\hat{\pi})(\psi_{s}^{(1)}\hat{\pi})\right]\right.\nonumber \\
 & +\left.E_{s}^{(1)}\frac{\partial}{\partial E_{s}^{(0)}}\left[\epsilon_{s}^{(0)}\frac{2}{\lambda_{s}^{(0)}}\right](\varphi_{s}^{*(0)}\hat{\varphi})(\varphi_{s}^{(0)}\hat{\varphi})+\epsilon_{s}^{(0)}\frac{2}{\lambda_{s}^{(0)}}\left[(\varphi_{s}^{*(1)}\hat{\varphi})(\varphi_{s}^{(0)}\hat{\varphi})+(\varphi_{s}^{*(0)}\hat{\varphi})(\varphi_{s}^{(1)}\hat{\varphi})\right]\right\} .\label{eq:deltaHe perturb}
\end{align}
Note that
\begin{align*}
\frac{\partial}{\partial E_{s}^{(0)}}\left[\epsilon_{s}^{(0)}\frac{\lambda_{s}^{(0)}}{2}\right] & =-\frac{1}{2E_{s}^{(0)}}\left[\epsilon_{s}^{(0)}\frac{\lambda_{s}^{(0)}}{2}-\frac{1}{1-E_{s}^{(0)}/4}\right]\underset{s\to0}{\longrightarrow}\frac{1}{12},\\
\frac{\partial}{\partial E_{s}^{(0)}}\left[\epsilon_{s}^{(0)}\frac{2}{\lambda_{s}^{(0)}}\right] & =\frac{1}{2E_{s}^{(0)}}\left[\epsilon_{s}^{(0)}\frac{2}{\lambda_{s}^{(0)}}-\frac{E_{s}^{(0)}}{1-E_{s}^{(0)}/4}\right]\underset{s\to0}{\longrightarrow}1,
\end{align*}
\end{widetext}and the limits of $\epsilon_{s}^{(0)}\lambda_{s}^{(0)}/2$
and $\epsilon_{s}^{(0)}2/\lambda_{s}^{(0)}$ when $s\to0$ are equal
to $1$ and $0$, respectively.

After substituting the expressions for $E_{s}^{(1)}$, $\psi_{s}^{(1)}$,
and $\varphi_{s}^{(1)}$ from Eqs.~(\ref{eq:Es})-(\ref{eq:phispert})
into Eq.~(\ref{eq:deltaHe perturb}), one obtains the results (\ref{eq:deltaHphi})
and (\ref{eq:deltaHpi}) from the main text where the superscript
$(0)$ was omitted.

\bibliographystyle{apsrev4-2}
\bibliography{DM_and_EH}

\end{document}